\documentstyle[12pt]{article}
\begin{document}
\title{Further results on Functional Determinants\\
 of Laplacians
in Simplicial Complexes}
\author{Erik Aurell $^1$, Per Salomonson $^2$\\\\
$^1$ Department of Mathematics,
Stockholm University \\
S--106 91 Stockholm,
Sweden \\
$^2$ Institute of Theoretical Physics, Chalmers University of Technology\\
S--412 96 G\"oteborg, Sweden}
\maketitle

\begin{abstract}
This paper is a sequel to a previous report
(Aurell~E. \& Salomonson~P., 1993), where we investigate the
functional determinant of the laplacian on piece-wise
flat two-dimensional surfaces, with conical singularities
in the interior and/or corners on the boundary.
Our results extend earlier
investigations of the determinants on
smooth surfaces with smooth boundaries.
The differences to the smooth
case are: a) different ``interaction energies''
between  pairs of conical singularities than one would
expect from a naive extrapolation of the results for
a smooth surface; and b) ``self-energies'' of the
singularities.
In this paper we give the results for general simplicial
complexes that are conformally
related. Special attention is given to the case of
disc topology with corners in the interior, and to the topology
of a sphere, where we can compare with alternative computations
in special cases where the spectra are known.
We consider both Dirichlet and
Neumann boundary conditions.
In the limit when all corners are almost flat
we recover the expressions for smooth surfaces with smooth
boundaries.
\end{abstract}

\bigskip
\centerline{\it Preprint, May 1994, hep-th/9405140 }
\newpage

\section{Introduction}
Gaussian integrals appear in many branches
in mathematical physics.
When $A$ is a symmetric real
finite-dimensional matrix with positive
eigenvalues, we have the elementary result
\begin{equation}
\label{gaussint}
\int \prod_{k=1}^n {{dx_k}\over{\sqrt{2\pi}}}
\, \exp(-{1\over 2} (\vec x, A\vec x)) = (\det A)^{-{1\over 2}}.
\end{equation}
If $A$ is an operator like the laplacian,
both sides of (\ref{gaussint}) vanish.
A regularization of the functional determinant
of an operator attempts
to give a meaning to (\ref{gaussint})
when the product of the eigenvalues
of $A$ diverges.
We refer to \cite{Parisi,Schwarzbook} for modern
discussions on regularizations of functional integrals.
One regularization proceeds from the
zeta function of the operator
\begin{equation}
\label{defZ}
Z_{A}(s) =\sum_{\nu:\lambda_{\nu}>0}\,  \lambda_{\nu}^{-s}.
\end{equation}
The sum in (\ref{defZ}) goes over the positive
eigenvalues of the operator, since we want to treat
operators having zero eigenvalues, and $s$ is a complex
number with sufficiently large real part so that the
sum converges.
Then, motivated by the formal result
${{dZ_{A}(s)}\over{ds}}|_{s=0} \sim -\sum_n\log\lambda_n$,
one defines
\begin{equation}
\label{defregdet}
\hbox{Regularized}(\det A) = \exp(-{{dZ_{A}(s)}\over{ds}}|_{s=0}).
\end{equation}
We study here functional determinants of scalar laplacians
in two-dimensional domains that are piece-wise flat
with isolated conical singularities, and boundaries that
are piece-wise straight with isolated corners.
We refer to these surfaces as simplicial complexes.

Functional determinants of the scalar two-dimensional laplacian
on {\it smooth} surfaces have been widely
studied\cite{Polyakov1,Alvarez,Schwarzbook}.
They appear in Polyakov's approach to string theory as
a double functional integral over embeddings in external space and
metrics in internal space.
The integral over embeddings is then just a Gaussian
integral like (\ref{gaussint}) with $A$ the scalar Laplace-Beltrami
operator. The differences in functional determinants between
two surfaces related by a conformal distortion can
be written
\begin{eqnarray}
\label{Polyakovaction}
Z'_{D'}(0) =
Z'_{D}(0) \mp {1\over{4\pi}}
\int_{\partial D} d\hat s \hat n\cdot\partial\sigma
+ {1\over{6\pi}}
\int_{\partial D} d\hat s \hat k\sigma \nonumber \\
+{1\over{12\pi}}
\int_{D} d^2z
\sqrt{\hat g} [\hat g^{ab} \partial_{a} \sigma
\partial_{b} \sigma  + \hat R\sigma].
\end{eqnarray}
In (\ref{Polyakovaction})
$\hat R$ is the Gaussian curvature of the reference
surface $D$, $\hat k$ is the geodetic curvature of the
boundary, and $\sigma =\log|{{dz'}\over{dz}}|$
(in local coordinates) is the conformal factor
\cite{Polyakov1,Alvarez,Weisberger}.
The $-$ ($+$) sign refer to Dirichlet (Neumann)
boundary conditions.

The kinetic energy term in
(\ref{Polyakovaction}) is infinite
for a mapping that changes the opening angle of a conical
singularity, although the quantity $Z_D'(0)$ is known
to be finite for many special cases (e.g. the equilateral
triangle or any rectangle,
see I or appendix~\ref{asolvable} below).
This kind of divergence
looks at first sight similar to electrostatics.
The interaction energy of a collection of charges can
be computed by multiplying a charge with the potential
from the other charges, and then summing over the charges.
One cannot include the potential from a charge on itself since
that would give an infinite and meaningless result.
One may also compute the interaction
energy by integrating the square of the electric field
over space, avoiding small
regions around each charge
(we assume for simplicity that the collection
has zero total charge so that the integrals converge
at infinity). That answer will then be divergent
with the radii of the small excluded volumes, but
if this divergence
is discarded one gets the right finite result.

One might therefore assume that in the case of
conical singularities,
(\ref{Polyakovaction}) is analogous to the
integral over the square of the
electric field,
and should be regularized by taking
away small regions around each singularity.
Our initial motivation for considering simplicial complexes
in detail was that this procedure does not work.
The answer is nevertheless sufficiently close for the
electrostatic picture to be useful.
There will be
a sum over interaction energies
between charges, but the potential
only reduces to the form of the standard electrostatic
potential in the limit when all the conical singularities are almost
flat. There will also be a finite
self-energy determined locally at each singularity.

This paper is a sequel to \cite{AurellSalomonson}, in the
following referred to as I,
where we investigated the functional determinants
in simplicial complexes with disc toplogy and corners
on the boundary, but without conical
singularities in the interior, i.e. polygons.
We extend here the results to general simplicial complexes
that can be related by conformal distortions.
Special attention is given to disc topology with
conical singularities in the interior,
and to spherical topology.
In these cases there are a number of special
integrable surfaces
where the spectra can be determined exactly,
and the functional determinants computed with
methods from number theory. These special results
fit our general expressions in all cases.

The paper is organized as follows: in section~\ref{s:hk}
we summarize the approach in I and the results obtained
there. The modifications necessary
when dealing with less simple topologies are included.
In sections~\ref{s:discDirichlet}
and~\ref{s:discNeumann} we treat simplicial complexes
of disc topology with Dirichlet and Neumann boundary
conditions. In section~\ref{s:sphere} we treat
simplicial complexes with the topology of a sphere.
In section~\ref{s:general} we consider the general case, albeit
in a less explicit form than for disc and spherical topology.
In section~\ref{s:comp} we show by an explicit example
that it is not possible to obtain the functional
determinants of a surface with conical singularities
by rounding off the corners and extrapolating to the
limit when the radius of the smoothing goes to zero.
In section~\ref{s:discussion}
we discuss the extension to more general variations,
and possible application to the original problem of the
Polyakov action.
In appendix~\ref{aexpan} we summarize relevant results from I.
In appendix~\ref{asolvable}
we list the
surfaces obtained
by quoting a torus with a finite order symmetry.
\section {Heat kernel and variational formula}
\label{s:hk}
We use the convention that the laplacian of a flat surface is
is $-(\partial^2_1+\partial^2_2)$ such that the spectrum
is non-negative.
The heat kernel for the laplacian on a domain $D$, and its trace,
can then be expressed
in terms of its eigenvalues and normalised eigenfunctions as
\begin{equation}
\label{heatkernel}
K_D (x,y,t) = \theta (t) \big{<}x\big{|} e^{-\Delta t} \big{|}y\big{>}
= \theta (t)\sum_{\nu}\Psi_{\nu}^*(x)\Psi_{\nu}(y) e^{-\lambda_{\nu}t},
\end{equation}
\begin{equation}
K_D (t) =\theta (t){\rm Tr}(e^{-\Delta t})
=\theta (t)\sum_{\nu}e^{-\lambda_\nu t}.
\end{equation}
As is well known\cite{Kac,McKean,Schwarzbook},
the trace of the heat kernel admits an expansion
for short times, that for two-dimensional surfaces with
a boundary goes as
\begin{equation}
 K_D(t) \sim {{c_1 }\over t} +
{{c_{1/2} }\over {t^{1\over 2}}} +
c_0 + {\cal O}(t^{1\over 2}) + \ldots
\end{equation}

The zeta function of the operator is the
Mellin transform of the
trace of the heat kernel, with the zero modes
subtracted:
\begin{equation}
\label{var3}
Z_D (s) = \sum_{\nu:\,\lambda_{\nu}>0} \lambda_{\nu}^{-s}= {1\over{\Gamma(s)}}
\int_0 ^\infty dt\, t^{s-1}\, [
{\rm Tr} (e^{-\Delta t}) -\dim\ker\Delta].
\end{equation}
The integral in (\ref{var3}) converges from any
finite time $t$ to infinity.
A term $t^{-\gamma}$ in the asymptotic expansion
for the heat kernel leads potentially to a pole
in the zeta function at $s=\gamma$ with residue
$c_{\gamma}/\Gamma(-\gamma)$. The exceptions
are zero and the negative integers, where the prefactor
$1/\Gamma(s)$ has a zero. Hence we have, e.g.,
\begin{equation}
 Z_D(0) = c_0 -\dim\ker\Delta.
\end{equation}
One may continue (\ref{var3}) analytically around
the poles at $s=1$ and $s=1/2$ to obtain for the derivative
at the origin:
\begin{eqnarray}
\label{zetaprimdef}
Z_D'(0) &=&
(3+\gamma)(c_0 -\dim\ker\Delta) - \nonumber \\
&&2\int_0^{\infty} dt \log t
{{\partial}\over{\partial t}}[
t^{1\over 2}{{\partial}\over{\partial t}}[t^{1\over 2}
{{\partial}\over{\partial t}} t (K_D(t)-\dim\ker\Delta) ]]
\end{eqnarray}
An alternative formula is obtained by taking
the integral from
$\epsilon$ to infinity, where $\epsilon$ is some small positive
number.
The integrand in (\ref{zetaprimdef})
behaves as $t^{-{1\over 2}}\log t $
for small $t$ so the limit
as $\epsilon$ goes to zero is harmless. Partial integrations
will however bring in divergent terms from the lower
boundary. A short calculation gives
\begin{equation}
\label{Alvarezcorr}
Z_D'(0) =
\gamma (c_0-\dim\ker\Delta)
+ \hbox{Finite}_{\epsilon\to 0}
\int_{\epsilon}^{\infty} {{dt}\over t} (K_D(t)-\dim\ker\Delta)
\end{equation}
where we understand that we keep the $\epsilon$-independent
part of (\ref{Alvarezcorr}), and throw away the
terms that diverge as $\epsilon$ tends to zero.
Equations (\ref{zetaprimdef}) and
(\ref{Alvarezcorr}) show that the derivative
of the zeta function at the origin exists as a well-defined
functional of the heat kernel, but are not of much practical
interest,
since they depend on the heat kernel
for large times in a way that is difficult to handle.

Things are better if one considers
variations of the laplacian.
Then (\ref{Alvarezcorr}) turns into
\begin{equation}
\label{generalvariation}
\delta [ Z_D'(0)] =
\gamma [\delta c_0)]
-\hbox{Finite}_{\epsilon\to 0}
\sum_{\nu}\,^{\prime}\,(\delta\lambda_{\nu})
e^{-\epsilon\lambda_{\nu}} {1\over{\lambda_{\nu} } }\, ,
\end{equation}
where the prime on the sum indicates that we sum only over the
non-zero eigenvalues.
The operators $e^{-\epsilon\Delta}$ and ${1\over{\Delta}}$ are
diagonal in the basis of eigenfunctions of $\Delta$. This allows
us to rewrite the sum in (\ref{generalvariation}) as
\begin{equation}
\label{bigsumandintegral}
\sum_{\nu}\,^{\prime}\,\sum_{\mu}\sum_{\kappa}
\int d^2 x \int d^2 y \int d^2 z
(\Psi_{\nu}^{*}(x) \delta\lambda_{\mu}\Psi_{\mu}(x))
(\Psi_{\mu}^{*}(y) e^{-\epsilon\lambda_{\kappa}} \Psi_{\kappa }(y))
(\Psi_{\kappa}^{*}(z) {1\over{\lambda_{\nu}}}\Psi_{\nu}(z))\, ,
\end{equation}
which can be rearranged to give
\begin{equation}
\label{bigsumandintegral2}
\delta [ Z_D'(0)] =
\gamma [\delta c_0)]
- \hbox{Finite}_{\epsilon\to 0}\int d^2 y \int d^2 z\,
[(\delta\Delta)_y G_{D}(y,z)]
K_D (z,y;\epsilon)\, .
\end{equation}
In (\ref{bigsumandintegral2}) $(\delta\Delta)_y$ is the variation
of the laplacian,
$K_D$ is the heat kernel, and
\begin{equation}
\label{Greensfunctiondef}
G_D(y,z) = \sum_{\nu}\,^{\prime}\,{1\over{\lambda_{\nu}}}
\Psi_{\nu}^{*}(y) \Psi_{\nu}(z)
\end{equation}
is the Green's function.
The heat kernel for short times only depend on the local metric
properties of the surface. The nonlocal information in
(\ref{bigsumandintegral2}) is therefore only contained in the
action of the varied laplacian on the Green's function.
We note that (\ref{bigsumandintegral2}) can be interpreted as a
regularization of the formal expression
$\delta(\log\det\Delta) = \hbox{Tr}[ (\delta\Delta)/\Delta]$.

In I and in this paper we only consider conformal variations.
In
conformal coordinates ($g_{ab}=e^{2\sigma (x)}\delta_{ab}$),
the laplacian is written
\begin{equation}
\Delta = -e^{-2\sigma (x)}(\partial_1^2 +\partial^2_2) .
\end{equation}
The action of the varied laplacian in (\ref{bigsumandintegral2})
then has the simple form
\begin{equation}
\label{GisinverseofD}
[(\delta\Delta)_y G_{D}(y,z)] = (-2\delta\sigma(y))\, [\, \delta^2 (y-z)
 -\sum_{\nu:\,\lambda_{\nu}=0} \Psi_{\nu}^{*}(y)\Psi_{\nu}(z)\, ] \, ,
\end{equation}
which gives
\begin{equation}
\label{var7}
\delta [ Z_D'(0)] =
\gamma [\delta c_0)]
+ \hbox{Finite}_{\epsilon\to 0} [
\int d^2 x\, 2\delta\sigma(x)K_D (x,x;\epsilon)
 -\sum_{\nu:\,\lambda_{\nu}=0}\int_D d^2x\, 2\delta\sigma(x)
|\Psi_{\nu}(x)|^2]\, .
\end{equation}

The scalar laplacian has in fact not more than one vanishing eigenvalue.
The corresponding normalized eigenfunction is $1/\sqrt{\hbox{Area}}$, and
(\ref{var7}) simplifies to
\begin{equation}
\label{var8}
\delta [ Z_D'(0)] =
\gamma [\delta c_0)]
+ \hbox{Finite}_{\epsilon\to 0} [
\int d^2 x\, 2\delta\sigma(x)K_D (x,x;\epsilon)
-{{\delta(\hbox{Area})}\over{\hbox{Area}}}]\, .
\end{equation}
On surfaces with a boundary and Dirichlet boundary conditions
the laplacian has no zero eigenvalue, and the correction
term in (\ref{var8}) is absent.

Our calculation is now brought into the form of
integrating a test
function ($\delta\sigma$) against the
heat kernel for short times, and then extracting
the time-independent piece.
One convenient aspect of simplicial complexes is that
there are qualitatively speaking only three different
kinds of short-time behaviours that need to be considered:
flat surface; close to a flat boundary, and close to a corner
or a conical singularity.
The flat surface does not contribute at all, and the
flat boundary gives a term that only depends on the topology
and may be absorbed in an undetermined constant of integration.

The entire contribution thus comes from the corners and
conical singularities.
The extraction of the finite piece in the short time
asymptotics was done in I, with a convention for
how the variations of the conical singularities
are parametrized.
Let us note that two conical singularities
with opening angles $\alpha$ and $\alpha+\delta\alpha$ can
be mapped from a flat surface by mappings that
in local coordinates ($z$ for the first singularity,
$z'$ for the second and $\omega$ for the flat surface) satisfy:
\begin{equation}
\label{loacalcoord}
{{dz}\over{d\omega}} = e^{\lambda}\omega^{\alpha-1}
\qquad {{dz'}\over{d\omega}} =
e^{\lambda+\delta\lambda}\omega^{\alpha+\delta\alpha-1}
\end{equation}
The variation of the
conformal factor of the mapping from $z$ to $z'$
then reads
\begin{equation}
\label{localcoorddelta}
\delta\sigma (z)
= \log|{{dz'}\over{dz}}| =
[\delta\lambda -{{\delta\alpha}\over{\alpha}}\lambda]
+{{\delta\alpha}\over{\alpha}}\log(\alpha z)
\end{equation}
The test function thus separates into one smooth
part, and one that is logarithmically divergent.

When the test function is smooth we see that we may take
its value at a corner outside of
(\ref{var7}).  The piece of the integral finite
as $\epsilon$ tends to zero will then give $Z_{\alpha}(0)$,
the contribution to $c_0$ or to $Z_D(0)$ ({\it not the derivative}),
from the corner. This contribution
has been computed long ago\cite{Kac,McKean,Dowker}:
it is ${1\over 24}({1\over{\alpha}}-\alpha)$ for a corner
on the boundary with opening angle
$\pi\alpha$, and ${1\over 12}({1\over{\alpha}}-\alpha)$
for a conical singularity
in the interior with opening angle $2\pi\alpha$.
We thus have one contribution to $\delta Z_D'(0)$
\begin{equation}
\label{loacalcoorddelta}
[\delta\lambda -{{\delta\alpha}\over{\alpha}}\lambda]
Z_{\alpha}(0)
\end{equation}
which we call nonlocal, as the scale factors $\lambda$
generally depend on the whole surface, and in particular
on the positions and strengths of the other conical
singularities.

The logarithmically divergent piece in (\ref{localcoorddelta}),
when inserted in (\ref{var8}) gives rise to a term
which only depends on the local
opening angle $\alpha$.
We can therefore
combine it with $\gamma c_0$,
the other term in (\ref{bigsumandintegral2}),
and write it as a total variation
$\delta Z_{\alpha}'(0)$. The integrated quantity
$Z_{\alpha}'(0)$ for a corner on the boundary with Dirichlet
boundary conditions was investigated in I.
The changes when considering a Neumann boundary conditions
or a conical singularity in the interior are small.
These results
are summarized in appendix~\ref{aexpan} below.

\section{Disc topology, Dirichlet boundary conditions}
\label{s:discDirichlet}
In this section we will establish convenient
representations of surfaces with the topology
of a disc, and then integrate
$Z'(0)$ using the formulae in
section~\ref{s:hk}.

Let us say loosely that we map the unit
disc (${\cal D}$, coordinate $u$) to ${\cal M}$
(coordinate $z$) by a transformation that satisfies
\begin{equation}
\label{sc1}
{{dz(u)}\over{du}}=e^{\lambda_0}
\prod_{u_{\mu} \in\, \partial {\cal D}}(u-u_{\mu})^{-\beta_{\mu}}
\prod_{u_i \in\, \hbox{int}\, {\cal D}}(u-u_i)^{-\beta_{i}}
(\bar u_i u-1)^{-\beta_{i}}
\end{equation}
which generalizes the Schwarz-Christoffel transform,
corners on the boundary at positions $z(u_{\mu})$,
to the case with conical singularities in the interior at positions
$z(u_{i})$. An opening angle at the boundary is written
$\pi\alpha_{\mu} = \pi(1-\beta_{\mu})$, and an opening
angle in the interior is written
$2\pi\alpha_i = 2\pi(1-\beta_i)$.
The image charges at positions $1/u_i$
serve to make the boundary of ${\cal M}$
piece-wise straight.
We will from now on use the convention that latin indices
($i,j,\ldots$) refer to conical singularities in the interior and their
image charges, while greek indices ($\mu,\nu,\ldots$)
refer to corners on the boundary.
The exterior angles must satisfy $\sum_i2\beta_i+\sum_{\mu}\beta_{\mu} =2$.

It is clear that
in general $z$ in (\ref{sc1}) is a coordinate defined
on a branched surface over a part of the complex plane
with cut lines properly identified.
The simplicial complex ${\cal M}$ can be pictured as paper triangles glued
together
at the sides. One of the triangles can be put down on the
plane. To construct the branched surface, one should
put down the other triangles on the plane, and when doing so, one has
to sever the triangles at the joints, if they make up less or
more than a full turn around a vertex.
Sides of triangles that have been unglued
should be identified. With this understanding one can take
$z$ in (\ref{sc1}) as a local variable everywhere on ${\cal M}$
and let (\ref{sc1}) define the conformal factor
$\sigma(u) = \log |{{dz(u)}\over{du}}|$.
The normalised area of ${\cal M}$ is chosen
\begin{equation}
\label{area1}
A({\cal M}) = \int_{\cal D} d^2u\,
\prod_{{\cal M}} |u-u_{\mu}|^{-2\beta_{\mu}}
|u-u_i|^{-2\beta_{i}}
|\bar u_i u-1|^{-2\beta_{i}}
\end{equation}
such that the real and the normalized areas are related by
\begin{equation}
\label{area2}
\hbox{Area} = e^{2\lambda_0}A({\cal M})
\end{equation}
It is a convenient fact that we may use (\ref{sc1}) to
smoothly interpolate between one simplicial complex ${\cal M}$ at
$a=0$, to another one ${\cal M}'$ at $a=1$ by
\begin{eqnarray}
\label{sc2}
{{dz(u,a)}\over{du}}&=e^{(1-a)\lambda_0+a\lambda_0'}
&\prod_{{\cal M}}(u-u_{\mu})^{-(1-a)\beta_{\mu}}
(u-u_i)^{-(1-a)\beta_{i}}
(\bar u_i u-1)^{-(1-a)\beta_{i}} \nonumber \\
&&\prod_{{\cal M}'}(u-u_{\mu'})^{-a\beta_{\mu'}}
(u-u_{i'})^{-a\beta_{i'}}
(\bar u_{i'} u-1)^{-a\beta_{i'}}
\end{eqnarray}
Each intermediate figure is then again a simplicial complex
with the same topology.
We may therefore without restriction consider variations
of $\lambda_0$ and the $\beta$'s, but leaving the
branchpoints $u_{\mu}$ and $u_i$ fixed.

We now separate the conformal factor at a corner or a conical
singularity into a local and a nonlocal part:
\begin{eqnarray}
\label{lambdas}
\sigma(u\sim u_{\mu}) &=& -\beta_{\mu}\log|u-u_{\mu}| + \lambda_{\mu}
+ {\cal O}(u-u_{\mu})\\
\lambda_{\mu} &=& \lambda_0 -
\sum_{\nu \neq\mu}\beta_{\nu}\log |u_{\mu}-u_{\nu}| -
\sum_{j}\beta_{j}\log |u_{\mu}-u_j| -
\sum_{j}\beta_{j}\log |\bar u_ju_{\mu}-1|\nonumber \\
&& \nonumber\\
\sigma(u\sim u_{i}) &=& -\beta_i\log|u-u_{i}| + \lambda_{i}
+ {\cal O}(u-u_{i}) \\
\lambda_{i} &=& \lambda_0 -
\sum_{\nu}\beta_{\nu}\log |u_{i}-u_{\nu}| -
\sum_{j\neq i}\beta_{j}\log |u_{i}-u_j| -
\sum_{j}\beta_{j}\log |\bar u_ju_{i}-1| \nonumber
\end{eqnarray}
We consider a variation of the simplex by varying
$\lambda_0$ and the $\beta$'s. Locally, that means that
we vary the opening angles $\beta_i$ and $\beta_{\mu}$
and the scale factors $\lambda_i$ and $\lambda_{\mu}$.
We then have the variation of the conformal factor in the
form of (\ref{localcoorddelta}), and we can directly write
\begin{eqnarray}
\label{deltaZ}
\delta Z_{{\cal M}}'(0)
&=& \delta [ \sum_i (2Z_{1-\beta_i}'(0) + {1\over 2}\log(1-\beta_i)) +
\sum_{\mu} Z_{1-\beta_{\mu}}'(0)]
+\nonumber \\
&& \sum_{\mu}({1\over{12}}({1\over{\alpha_{\mu}}}-\alpha_{\mu}))
[\delta\lambda_{\mu}-
{{\delta\alpha_{\mu}}\over{\alpha_{\mu}}}\lambda_{\mu}]
 +\nonumber \\
&& \sum_{i}({1\over{6}}({1\over{\alpha_{i}}}-\alpha_{i}))
[\delta\lambda_{i}-
{{\delta\alpha_{i}}\over{\alpha_{i}}}\lambda_{i}]
\end{eqnarray}
We notice that the nonlocal quantities in (\ref{deltaZ})
may be rewritten as
\begin{equation}
\label{deltaZnonlocal1}
\delta[{1\over{12}}\sum_{\mu}
({1\over{\alpha_{\mu}}}-\alpha_{\mu})\lambda_{\mu}+
{1\over{6}}\sum_{i}
({1\over{\alpha_{i}}}-\alpha_{i})\lambda_{i}]
+ {1\over 6}\sum_{\mu}\delta\alpha_{\mu}\lambda_{\mu}
+ {1\over 3}\sum_{i}\delta\alpha_{i}\lambda_{i} .
\end{equation}
and that the extra terms in (\ref{deltaZnonlocal1})
may be integrated to
\begin{eqnarray}
\label{deltaZnonlocal2}
&&\sum_{\mu} ( {1\over{12}}\sum_{\nu\neq\mu}\beta_{\mu}\beta_{\nu}
\log|u_{\mu}-u_{\nu}|
+ {1\over{3}}\sum_{j}\beta_{\mu}\beta_{j}
\log|u_{\mu}-u_{j}| ) +\nonumber \\
&&\sum_{i} ( {1\over{6}}\sum_{j\neq i}\beta_{i}\beta_{j}
\log|u_{i}-u_{j}|
+ {1\over{6}}\sum_{j}\beta_{i}\beta_{j}
\log|\bar u_{j}u_{i}-1| )
\end{eqnarray}
Now we can combine (\ref{deltaZ}), (\ref{deltaZnonlocal1})
and (\ref{deltaZnonlocal2}) and integrate the
functional determinant of the Laplacian
on a simplicial complex with disc topology and Dirichlet
boundary conditions to:
\begin{eqnarray}
\label{integratedZ}
Z_{{\cal M},\hbox{Area}}'(0)
&=& \sum_i (2Z_{1-\beta_i}'(0) +
{1\over 2}\log(1-\beta_{i}))+
\sum_{\mu}Z_{1-\beta_{\mu}}'(0) +
Z_{{\cal M}}(0)\log{{\hbox{Area}}\over{A({\cal M})}} \nonumber \\
&& {-{1\over 12}}\sum_{\mu} {{\beta_{\mu}}\over{1-\beta_{\mu}}}
[\sum_{\nu\neq\mu}
\beta_{\nu}
\log|u_{\mu}-u_{\nu}|
+ \sum_{j}\beta_{j}
\log|u_{\mu}-u_{j}| \nonumber \\
&& \qquad + \sum_{j}\beta_{j}
\log|\bar u_{j} u_{\mu}-1|\,  ] +\nonumber \\
&& {-{1\over 6}}\sum_{i}{{\beta_{i}}\over{1-\beta_{i}}}[
\sum_{j\neq i}\beta_{j}
\log|u_{i}-u_{j}|
+\sum_{\nu}\beta_{\nu}
\log|u_{i}-u_{\nu}| \nonumber \\
&& \qquad + \sum_{j}\beta_{j}
\log|\bar u_{j} u_{i}-1|\, ] \nonumber \\
&& \qquad +\,\hbox{Integration constant}
\quad\hbox{(Disc topology, Dirichlet b.c.)}
\end{eqnarray}
The result (\ref{integratedZ}) does not change if the
branchpoints $u_{\mu}$ and $u_i$ are moved around by a Moebius
transformation that leaves the unit disc invariant,
although the normal area and the corner-corner interaction
terms vary when taken separately.
Formula (\ref{integratedZ}) holds with or without conical
singularities in the interior.
In I the integration constant
was determined to be zero.

In appendix~\ref{asolvable} one can find two one-parameter
families and seven more symmetric special surfaces of
disc topology, for which the spectral functions can
be computed directly. It can be checked that the general
formula (\ref{integratedZ}) reproduces the exact results in all
cases.

In the limit where all exterior angles are small we have
\begin{eqnarray}
\label{integratedZdiscapprox}
Z_{{\cal M},\hbox{Area}}'(0)
&\sim& -2{{dZ_{\alpha}'(0)}\over{d\alpha}}|_{\alpha=1}
- {1\over 2}\sum_i\beta_i
+{1\over 6}\log{{\hbox{Area}}\over{A({\cal M})}} \nonumber \\
&& -{1\over{12}}\sum_{\mu}\beta_{\mu} [ \sum_{\nu\neq\mu}
\beta_{\nu}
\log|u_{\mu}-u_{\nu}|+
\sum_{j}\beta_{j}
\log|u_{\mu}-u_{j}| \nonumber \\
&& \qquad + \sum_{j}\beta_{j}
\log|\bar u_{j} u_{\mu}-1| \, ] \nonumber \\
&& -{1\over{6}}\sum_{i}\beta_{i}[ \sum_{j\neq i}\beta_{j}
\log|u_{i}-u_{j}|
+\sum_{\nu}\beta_{\nu}
\log|u_{i}-u_{\nu}| \nonumber \\
&& \qquad + \sum_{j}\beta_{j}
\log|\bar u_{j} u_{i}-1|\, ] \nonumber \\
&& + \, \hbox{Integration constant}
\end{eqnarray}
Comparing (\ref{zprimexpb})
with the exact result for the functional determinant
on a circular disc\cite{Weisberger} we have
\begin{eqnarray}
\label{exactdisc}
-2{{dZ_{\alpha}'(0)}\over{d\alpha}}|_{\alpha=1}
&=& Z_{\hbox{disc, Dirichlet b.c.}}'(0)\nonumber \\
& =& {1\over 6}\log 2 + {1\over 2}\log\pi + {5\over 12}
+ 2\zeta'(-1)
\end{eqnarray}

We may enclose every branchpoint $u_i$ in the interior
of the disc in a small
circular curve $C_i$ with radius $r_i$,
and every branchpoint $u_{\mu}$ on the boundary in a
small semicircle $C_{\mu}$.
We may then approximate the interaction terms in
(\ref{integratedZdiscapprox}) as
\begin{eqnarray}
-{1\over {12\pi}}\int_{C_i,C_{\mu}}
[\hat n\cdot\partial\sigma](\sigma -\lambda_0) ds
+ \sum_i{1\over{6}}\beta_i^2\log r_i
+\sum_{\mu}{1\over{12}}\beta_{\mu}^2\log r_{\mu}  + {\cal O}(r_i,r_{\mu})
\end{eqnarray}
where $\hat n$ are the normals directed {\it away from}
the points $u_i$ and $u_{\mu}$.
We may close off the integration by including the circular
segments between the branchpoints $u_{\mu}$.
The extra integrals are evaluated by noting
that by the Cauchy-Riemann equations
\begin{equation}
\label{CauchyRiemanneqs}
\hat n\cdot\partial\sigma|_{|u|=1} = - {d\over{d\phi}}\hbox{Im}\log
{{dz}\over{du}}|_{u=e^{i\phi}}
\end{equation}
and the right-hand of (\ref{CauchyRiemanneqs}) is
equal to $1$ between the branchpoints.
Then take the $\beta_i$'s, $\beta_{\mu}$'s, $r_i$'s and the $r_{\mu}$'s
to zero such that correction terms
vanishes, and after an integration by parts
one finds:
\begin{eqnarray}
\label{Liouville2disc}
{1\over {12\pi}}\int_{{\cal D}}
(\partial\sigma)^2 d^2u +
{1\over {12\pi}}\int_{\partial{\cal D}}
(\sigma-\lambda_0) ds
\end{eqnarray}
The line integral in (\ref{Liouville2disc}) vanishes.
We may identify
\begin{eqnarray}
\label{Liouville3discextra}
{1\over 3}\lambda_0 = {1\over {6\pi}}\int_{\partial{\cal D}}
\sigma ds \qquad\hbox{and}\qquad
-{1\over 2}\sum_i\beta_i = -{1\over {4\pi}}\int_{\partial{\cal D}}
(\hat n\cdot\partial\sigma) ds.
\end{eqnarray}
Including everything from (\ref{integratedZdiscapprox}),
(\ref{exactdisc}), (\ref{Liouville2disc}),
(\ref{Liouville3discextra})
and writing the integrals as over the disc
with usual Cartesian
volume element written as $\sqrt{\hat g}d^2 z$,
and the line element on the circle written $d\hat s$,
we have
\begin{eqnarray}
\label{Liouvilleactiondisc}
Z_{{\cal M},\hbox{Area}}'(0)
&=& Z_{\hbox{disc}}'(0)
+{1\over{12\pi}}
\int_{{\cal D}} d^2z
\sqrt{\hat g} \hat g^{ab} \partial_{a} \sigma
\partial_{b} \sigma  \nonumber \\
&& + {1\over {6\pi}}\int_{\partial{\cal D}}
\sigma d\hat s
-{1\over {4\pi}}\int_{\partial{\cal D}}
(\hat n\cdot\partial\sigma) d\hat s
\nonumber \\
&&\nonumber \\
&&\qquad +\, \hbox{Integration constant}
\quad\hbox{(Disc topology, Dirichlet b.c.)}
\end{eqnarray}
Since the exact value of $Z_{\hbox{disc}}'(0)$
is included, we have determined again the integration
constant in (\ref{Liouvilleactiondisc}) to be zero.

\section{Disc topology, Neumann boundary conditions}
\label{s:discNeumann}

The changes when introducing Neumann boundary conditions
are small. The mapping from the unit disc and the
conformal factor are unchanged.
The nonlocal terms in (\ref{integratedZ})
are not sensitive to the kind of boundary conditions
imposed. The boundaries change the local self-energies, and
the zero-mode has to be deducted from diagonal element of the
heat kernel according to (\ref{var8}).

We thus have the
functional determinant of the Laplacian
on a simplicial complex with disc topology and Neumann
boundary conditions as:
\begin{eqnarray}
\label{integratedZNeumann}
Z_{{\cal M},\hbox{Area}}'(0)
&=& \sum_i(2Z_{1-\beta_i}'(0) + {1\over 2}\log(1-\beta_i))+\sum_{\mu}(
Z_{1-\beta_{\mu}}'(0)+
{1\over 2}\log(1-\beta_{\mu})) \nonumber \\
&& + Z_{{\cal M}}(0)\log{{\hbox{Area}}\over{A({\cal M})}} -\log\hbox{Area}
\nonumber \\
&& -{1\over{12}}\sum_{\mu} {{\beta_{\mu}}\over{1-\beta_{\mu}}}
[ \sum_{\nu\neq\mu}
\beta_{\nu}
\log|u_{\mu}-u_{\nu}|
+ \sum_{j}\beta_{j}
\log|u_{\mu}-u_{j}| \nonumber \\
&& \qquad + \sum_{j}\beta_{j}
\log|\bar u_{j} u_{\mu}-1|\,  ] \nonumber \\
&&-{1\over{6}} \sum_{i}{{\beta_{i}}\over{1-\beta_{i}}}[ \sum_{j\neq i}\beta_{j}
\log|u_{i}-u_{j}|
+\sum_{\nu}\beta_{\nu}
\log|u_{i}-u_{\nu}| \nonumber \\
&& \qquad + \sum_{j}\beta_{j}
\log|\bar u_{j} u_{i}-1|\, ] \nonumber \\
&& \qquad +\, \hbox{Integration constant}
\quad\hbox{(Disc topology, Neumann b. c.)}
\end{eqnarray}

Comparing
with the exact result for the functional determinant
on a circular disc\cite{Weisberger} we have
\begin{eqnarray}
\label{exactdiscNeumann}
-2{{dZ_{\alpha}'(0)}\over{d\alpha}}|_{\alpha=1}
&=& Z_{\hbox{disc,Neumann}}'(0)+(1+\log\pi)\nonumber \\
& =& ({1\over 6}\log 2 - {1\over 2}\log\pi - {7\over 12}
+ 2\zeta'(-1))+(1+\log\pi)
\end{eqnarray}
One sees that in the smooth limit
\begin{eqnarray}
{1\over 2}\sum_i\log(1-\beta_i) + {1\over 2}\sum_{\mu}\log(1-\beta_{\mu}) \sim
-1 + {1\over 2}\sum_i\beta_i
\end{eqnarray}
We can therefore write down the result
in analogy to (\ref{Liouvilleactiondisc}) as
\begin{eqnarray}
\label{LiouvilleactiondiscNeumann}
Z_{{\cal M},\hbox{Area}}'(0)
&=& Z_{\hbox{disc}}'(0)
+{1\over{12\pi}}
\int_{{\cal D}} d^2z
\sqrt{\hat g} \hat g^{ab} \partial_{a} \sigma
\partial_{b} \sigma  \nonumber \\
&& + {1\over {6\pi}}\int_{\partial{\cal D}}
\sigma d\hat s
+{1\over {4\pi}}\int_{\partial{\cal D}}
(\hat n\cdot\partial\sigma) d\hat s
-\log{{\hbox{Area}}\over{\pi}}\nonumber \\
&& \nonumber \\
&&\qquad +\, \hbox{Integration constant}
\quad\hbox{(Disc topology, Neumann b.c.)}
\end{eqnarray}
Since the exact value of $Z_{\hbox{disc}}'(0)$
is included, we have determined the integration
constant in (\ref{LiouvilleactiondiscNeumann}) to be zero.

\section{Spherical topology}
\label{s:sphere}
In this section we will establish convenient
representations of surfaces with the topology
of a sphere, and then integrate
$Z'(0)$ using the formulae in
section~\ref{s:hk}.
Visualize the
sphere with unit radius as a ball lying on the plane.
Take the point where the plane and the sphere touch to be
the south pole of the sphere and the origin in the plane.
By a stereographic projection,
$r=2\tan {{\pi-\theta}\over 2}$, we map the point
$(\theta,\phi)$ in spherical
coordinates on the sphere to $(r,\phi)$ in polar
coordinates in the
plane.
The volume element on the sphere, $\sin\theta d\phi d\theta$,
is in polar coordiantes  ${1\over{(1+(r/2)^2)^2}}rd\phi dr$.

Let the flat laplacian in the plane be $\hat\Delta $
and the laplacian on the sphere $\tilde\Delta$.
Writing the laplacian on the sphere in coordinates
$r$ and $\phi$, it is then related to the flat laplacian
by
\begin{equation}
\tilde\Delta = (1+(r/2)^2)^2\hat\Delta
\end{equation}
Call the conformal factor of the stereographic projection from
the plane to the sphere $\tilde\sigma$. We have
\begin{equation}
\tilde R = e^{-2\tilde\sigma}(\hat R + 2 \hat\Delta\tilde\sigma)
\end{equation}
where $\tilde R$ (equal to $2$) is the Gaussian curvature
of a point on the sphere and $\hat R$ (equal to zero)
is the curvature of the plane.
We can now map the plane
(coordinate $\omega$)
to a simplicial complex with the topology of a sphere
(${\cal M}$, coordinate $z$)
by a transformation that satisfies
\begin{equation}
\label{scSphere}
{{dz(\omega)}\over{d\omega}}=e^{\lambda_0}
\prod_{i}(\omega-\omega_i)^{-\beta_i}
\end{equation}
The exterior angles must satisfy $\sum_i 2\beta_i = 4$.
The normalised area of ${\cal M}$ is chosen
\begin{equation}
\label{areaSphere1}
A({\cal M}) = \int_{\bar C} d^2\omega\,
\prod_{{\cal M}} |\omega -\omega_i|^{-2\beta_i}
\end{equation}
such that the real and the normalized areas are related by
\begin{equation}
\label{areaSphere2}
\hbox{Area} = e^{2\lambda_0}A({\cal M})
\end{equation}
We will call the conformal factor defined by (\ref{scSphere})
$\hat\sigma$.
By combining (\ref{scSphere}) with a stereographic projection,
we obtain a map from the sphere to ${\cal M}$, for
which the conformal factor is
\begin{equation}
\sigma = {1\over 2}
\log|{{\partial (z,\bar z)}\over{\partial(\theta,\phi)}}|
= \hat\sigma - \tilde\sigma
\end{equation}

We may now proceed in analogy with the case of
a simplicial complex with topology of a disc,
and write down the functional determinant
for a simplicial complex of spherical topology:
\begin{eqnarray}
\label{integratedZSphere}
Z_{{\cal M},\hbox{Area}}'(0)
&=& \sum_i (2Z_{1-\beta_i}'(0)+{1\over 2}\log(1-\beta_i)) +Z_{{\cal
M}}(0)\log{{\hbox{Area}}\over{A({\cal M})}} \nonumber \\
&&-{1\over{6}}\sum_{i,\, j\neq i}{{\beta_{i}\beta_{j}}\over{1-\beta_{i}}}
\log|\omega_{i}-\omega_{j}| -\log\hbox{Area} \nonumber \\
&& \qquad  +\,  \hbox{Integration constant}
\quad\hbox{(Spherical topology)}
\end{eqnarray}
We determine below the integration constant
in (\ref{integratedZSphere}) to be $\log 2$.

In appendix~\ref{asolvable} one can find one two-parameter
family and five more symmetric special surfaces of
disc topology, for which the spectral functions can
be computed directly. It can be checked that the general
formula (\ref{integratedZSphere}) reproduces the exact results in all
cases.

In the limit where all exterior angles are small we have
\begin{eqnarray}
\label{integratedZSphereapprox}
Z_{{\cal M},\hbox{Area}}'(0)
&\sim& -4{{dZ_{\alpha}'(0)}\over{d\alpha}}|_{\alpha=1}-1+
{1\over 3}\log{{\hbox{Area}}\over{A({\cal M})}} -
{1\over{6}}\sum_{i,\, j\neq i}\beta_{i}\beta_{j}
\log|\omega_{i}-\omega_{j}| \nonumber \\
&& \qquad -\log\hbox{Area} +  \hbox{Integration constant}
\end{eqnarray}
Comparing (\ref{zprimexpb})
with the exact result for the functional determinant
on a sphere\cite{Weisberger}, we have
\begin{eqnarray}
\label{exactsphere}
-4{{dZ_{\alpha}'(0)}\over{d\alpha}}|_{\alpha=1} -1
= Z_{\hbox{sphere}}'(0) -{2\over 3}\log 2 +\log 2\pi + {1\over 3}
\end{eqnarray}
We may enclose every branchpoint $\omega_i$ in a small
circular curve $C_i$ with radius $r_i$ and approximate
\begin{eqnarray}
\label{integralapprox}
-{1\over{6}}\beta_i\sum_{\, j\neq i} \beta_{j}
\log|\omega_{i}-\omega_{j}| &=&
{1\over{6}}\beta_i(\hat\sigma-\lambda_0+\beta_i\log r_i)
+ {\cal O}(r_i)\nonumber \\
&=&
-{1\over {12\pi}}\int_{C_i}
[\hat n\cdot\partial\hat\sigma]\hat\sigma ds
-{1\over{6}}\lambda_0 + {1\over{6}}\beta_i^2\log r_i + {\cal O}(r_i)
\end{eqnarray}
where $\hat n$ is normal directed {\it away from}
the point $\omega_i$.
Closing off with one circle at $|\omega|=R$ far outside
all the $\omega_i$'s, taking
the $\beta_i$'s and the $r_i$'s
to zero such that ${1\over{6}}\sum_i\beta_i^2\log r_i$
vanishes
and integrating by parts, we find:
\begin{eqnarray}
\label{Liouville1}
-{2\over 3}\log R +{1\over {12\pi}}\int_{|\omega|<R}
(\partial\hat\sigma)^2 d^2\omega
+{\cal O}(1/R)
\end{eqnarray}
The integral in (\ref{Liouville1}) can
be rewritten as
\begin{eqnarray}
\label{Liouville2}
{1\over {12\pi}}\int_{|\omega|<R}
(\partial\sigma)^2+ 2 \partial\sigma\cdot\partial\tilde\sigma
+ (\partial\tilde\sigma)^2 d^2\omega
\end{eqnarray}
The integral over $(\partial\sigma)^2$ is convergent at infinity
and may be closed.
The second term may be rewritten as
\begin{eqnarray}
\label{Liouville3}
-{2\over 3}\lambda_0 + {4\over 3}\log 2
+
{1\over {12\pi}}\int_{|\omega|<R}
\sigma\tilde R e^{2\tilde\sigma} d^2\omega
\end{eqnarray}
The integral in (\ref{Liouville3}) is again convergent at
infinity and may be closed.
The third term in (\ref{Liouville2}) is simply
${2\over 3}\log R - {2\over 3}\log 2 -{1\over 3}$.

Including everything from (\ref{integratedZSphereapprox}),
(\ref{exactsphere}) and (\ref{Liouville1})
and writing the integrals as over the sphere
with usual
volume element written as $\sqrt{\tilde g}d^2 z$
we have
\begin{eqnarray}
\label{Liouvilleaction}
Z_{{\cal M},\hbox{Area}}'(0)
= Z_{\hbox{sphere}}'(0)
+{1\over{12\pi}}
\int_{S^2} d^2z
\sqrt{\tilde g} [\tilde g^{ab} \partial_{a} \sigma
\partial_{b} \sigma  + \tilde R\sigma] \nonumber \\
-\log\hbox{Area} + \log2\pi + \hbox{Integration constant}
\end{eqnarray}
Since the area of the unit sphere is $4\pi$,
the integration constant in (\ref{Liouvilleaction})
must be $\log 2$.

\section{The general simplicial complex}
\label{s:general}
When we extend extend the approach
to functional determinants through simplical approximations
to higher genus surfaces,
there is the problem of
Teichmuller parameters, which prevents
us from going between two arbitrary surfaces with a
conformal transformation.
There is also a lack of integrable cases to fix
an integration constant and use as checks on the calculations.
We are not aware of any two surfaces with genus one or
higher which are related by a conformal distortion
and for which both spectra of the laplacian are known.

Here we will sketch how to compute the functional
determinants under the restriction that all variations
are conformal.
For simplicity we will consider closed surfaces.
The cases with Dirichlet and Neumann boundary conditions
are easily obtained in analogy with sections~\ref{s:discDirichlet}
and~\ref{s:discNeumann}.
We may then say that we have two simplicial
complexes ${\cal M}_1$  and
${\cal M}_2$, such that the second
can be parametrized by the coordinates
of the first, with a metric $g$ which is
related to the piece-wise flat metric $\hat g$  on  ${\cal M}_1$
by $g_{ab}=e^{\sigma}\hat g_{ab}$.

The conformal factor $\sigma$ can be thought of as
an electrostatic potential of a collection of charges
on ${\cal M}_1$ as follows:
suppose that at the point labelled by coordinate
$z^{(1)}_i$ on ${\cal M}_1$
there is a conical singularity of ${\cal M}_1$,
or a conical singularity of ${\cal M}_2$, or both.
Then we can interpret the differences in exterior angles
as charges of strength
$(\beta_i^{(2)}-\beta_i^{(1)})$.
We note that the sum of these charges is zero.
The potential is determined up to an additive constant, which
is fixed by comparing the areas of
${\cal M}_1$ and ${\cal M}_2$.

Now we may interpolate linearly between ${\cal M}_1$
(at $a=0$) and ${\cal M}_2$ (at $a=1$) by considering the
family of electrostatic potentials generated by charges
$a\cdot(\beta_i^{(2)}-\beta_i^{(1)})$, and an overall
area changing term $a\cdot ( \lambda_0^{(2)}-\lambda_0^{(1)})$.
The electrostatic potential obeys an equation linear in the charges.
By changing all charges by an overall factor, we will necessarily
only change the electrostatic potential in the same proportion.
Let us therefore write
\begin{equation}
\label{sigmafunctionofa}
\sigma(z;a) = a\cdot G(z;\{ z_i^{(1)} \};
\{ (\beta_i^{(2)}-\beta_i^{(1)})\})
+ a\cdot(\lambda_0^{(2)}-\lambda_0^{(1)}),
\end{equation}
where $G$ is the electrostatic potential at the point $z$
from the full charges.
Close to a conical singularity we may separate out the
logarithmically divergent term of the variation, and write
\begin{equation}
\label{lambdafunctionofa}
\delta\sigma(z\sim z_i;a) \sim
{{\delta\alpha_i(a)}\over{\alpha_i(a)}}
\log |z-z_i| + [\delta\lambda_i(a) -
{{\delta\alpha_i(a)}\over{\alpha_i(a)}}\lambda_i(a)]
\end{equation}
where
\begin{equation}
\lambda_i(a) = a\cdot\lambda_i(\{ z_i^{(1)} \},
\{ (\beta_i^{(2)}-\beta_i^{(1)}) \})
+a\cdot(\lambda_0^{(2)}-\lambda_0^{(1)})
\end{equation}
contains the non-singular part of $G$.
The variation (\ref{lambdafunctionofa})
of the conformal factor can be integrated
against the heat kernel around the conical singularity
to give (\ref{deltaZ}), with only a different definition of
the $\lambda_i$'s.
That gives the same self-energies as previously.
The interactions between conical singularities
may be obtained as in (\ref{deltaZnonlocal1}) by noting
that both $\delta\alpha_i$  and $\lambda_i$ depend linearly
on $a$ and on the angle differences $(\beta_i^{(2)}-\beta_i^{(1)})$.
We can then write
write down the change of the functional
determinant when going from one simplicial
complex ${\cal M}_1$ with area $\hbox{Area}_1$ to another
simplicial complex ${\cal M}_2$ with area $\hbox{Area}_2$.
We write the normal area of ${\cal M}_2$, considered as a conformal
distortion of ${\cal M}_1$, as a function also of
${\cal M}_1$ and $\hbox{Area}_1$:

\begin{eqnarray}
\label{integratedgeneralZ}
Z_{{\cal M}_2,\hbox{Area}_2}'(0)
&= & Z_{{\cal M}_1,\hbox{Area}_1}'(0) + \sum_i (2Z_{1-\beta_i^{(2)}}'(0) +
{1\over 2}\log(1-\beta_{i}^{(2)})) \nonumber \\
&& -\sum_i (2Z_{1-\beta_i^{(1)}}'(0) +
{1\over 2}\log(1-\beta_{i}^{(1)}))
+ Z_{{\cal M}_2}(0)\log{ {\hbox{Area}_2} \over {A({\cal M}_2;
{\cal M}_1,\hbox{Area}_1)} }
\nonumber \\
& & -{1\over 6}\sum_{i}
({1\over{\alpha_{i}^{(2)}}} - \alpha_{i}^{(1)})
\lambda_{i}(\{ z_i^{(1)}\};\{(\beta_i^{(2)}-\beta_i^{(1)})\})
-\log{ {\hbox{Area}_2} \over {\hbox{Area}_1} }
\end{eqnarray}

In the limit when all the angle differences are small and all the
angles are close to one, (\ref{integratedgeneralZ}) tends to
\begin{eqnarray}
\label{integratedgeneralZapprox}
Z_{{\cal M}_2,\hbox{Area}_2}'(0)
&\sim& Z_{{\cal M}_1,\hbox{Area}_1}'(0) + {{\chi}\over
6}\log{{\hbox{Area}_{2}}\over{A({\cal M}_2;{\cal M}_1,\hbox{Area}_1)}}
\nonumber \\
&& -{1\over 6}\sum_{i}
((\beta_{i}^{(2)} - \beta_{i}^{(1)})+2\beta_i^{(1)})
\lambda_{i}(\{ z_i^{(1)}\};\{(\beta_i^{(2)}-\beta_i^{(1)})\}) \nonumber \\
&&-\log{ {\hbox{Area}_2} \over {\hbox{Area}_1} }
\end{eqnarray}
where $\chi$ is the Euler characteristic of the surfaces.

We can rewrite the interaction term
in (\ref{integratedgeneralZapprox}) as line integrals along small
circular curves around the branchpoints in the same way as
in  (\ref{integralapprox}) for spherical toplogy,
and we recover the general result (\ref{Polyakovaction}).
In addition we have the contribution from the
zero mode $-\log{ {\hbox{Area}_2} \over {\hbox{Area}_1} }$.

\section{Comparison with smooth limits}
\label{s:comp}

We will here consider what happens when one tries to
approximate the functional determinant of a simplicial
complex, with the one on a smooth surface obtained by rounding off
the corners with some characteristic radius $\epsilon$.

This leads to problems with both the corner self-energies and
the interaction energy. A smooth surface only feels the
linear term in the expansion of $Z_{\alpha}'(0)$ around
$\alpha=1$. The interaction energies have a caracteristic
feature ${{\beta_i\beta_j}\over{1-\beta_i}}$, which is indistinguishable
from $\beta_i\beta_j$ when all the $\beta$'s are small,
but different when they are not.

It is nice to look at a concrete example.
Triangles with Dirichlet boundary conditions can be
mapped from the unit disc with branchpoints that
we may put without loss of generality at
$u_{\mu}=e^{2\pi i\mu/3},$ $\mu=1,2,3$.
Then the interaction term simplifies, and we find:
\begin{equation}
\label{tri2}
Z_T'(0) =
\sum_{\mu=1,2,3} Z'_{\alpha_{\mu}}(0)+Z_T(0)\log{{\hbox{Area}}\over{A(T)}},
\end{equation}
if the normal area is chosen to be
\begin{equation}
\label{tri3}
A(T)=
{{\pi\Gamma(\alpha_1)\Gamma(\alpha_2)\Gamma(\alpha_{3})}\over
{2\Gamma(1-\alpha_1)\Gamma(1-\alpha_2)\Gamma(1-\alpha_{3})}}
\end{equation}
which corresponds to the particular choice $\lambda_0={1\over 2}\log 3$
(see I).
The relevant continuous action is (\ref{Liouvilleactiondisc}).
For two triangles we have two conformal factors
\begin{eqnarray}
\label{sigmatriangle}
\sigma_1(u) = \lambda_0 -\sum_{\mu} \beta_{\mu}^{(1)}
\log|u-u_{\mu}| \qquad
\sigma_2(u) = \lambda_0 -\sum_{\mu} \beta_{\mu}^{(2)}
\log|u-u_{\mu}|
\end{eqnarray}
The integral of the conformal factors over the boundary
is equal to $\lambda_0/3$, and does not differ between
two triangles.
The integral of the normal derivative at the boundary
is zero, because there are no singularities in the interior.
The difference of the the
kinetic energy integrals in (\ref{Liouvilleactiondisc})
can be written out as
\begin{eqnarray}
\sum_{\mu} {{(\beta_{\mu}^{(2)})^2-(\beta_{\mu}^{(1)})^2}\over{12\pi}}
\int_{{\cal D}} d^2u\,
{1\over{|u-u_{\mu}|^2}} +\nonumber \\
\sum_{\nu\neq\mu}
{{\beta_{\mu}^{(2)}\beta_{\nu}^{(2)}
-\beta_{\mu}^{(1)}\beta_{\nu}^{(1)}}\over{12\pi}}
\int_{{\cal D}} d^2u\,
{\rm Re} [{1\over{\overline{u-u_{\mu}}}}{1\over{u-u_{\nu}}}]
\end{eqnarray}
The diagonal terms are logarithmically divergent.
Integrating over the disc except for small semi-discs
of radius $\epsilon$ around each $u_{\mu}$ one
obtains
\begin{eqnarray}
\label{diagonalterms}
\sum_{\mu} {{(\beta_{\mu}^{(2)})^2-(\beta_{\mu}^{(1)})^2}\over{12}}
\log {1\over{\epsilon}} + {\cal O}(\epsilon)
\end{eqnarray}
The off-diagonal terms are finite and integrate to
\begin{eqnarray}
\label{offdiagonalterms}
\sum_{\mu} {{(\beta_{\mu}^{(2)})^2-(\beta_{\mu}^{(1)})^2}\over{12}}(1+\log 2),
\end{eqnarray}
if one eliminates the cross products using $\sum_{\mu}\beta_{\mu}=2$.
A comparison with (\ref{tri2}) and (\ref{cornerprime16},\ref{cornerprime17})
shows that the simple result of (\ref{offdiagonalterms}) cannot
give the complete difference of the functional determinants
between two triangles.
\section{Discussion}
\label{s:discussion}

We have in this paper investigated functional
determinants on simplicial complexes related by
conformal distortions.
The variation of the functional determinant is then
given by the short-time behaviour of the heat kernel,
weighted by the variation of the conformal factor
(\ref{var8}).

A general pair of simplicial
complexes of genus one or higher are not related by conformal
distortions. The methods of this paper can then only give
a partial answer.
It would therefore be worthwhile
to know how to compute the variations
under small distortions that are not conformal.
For simplicial complexes
the most obvious candidate is linear shear.
Even in the absence of integrable cases, that would
reduce the undetermined integration constants to one
for each topology, which is tolerable.
The general variational formula for such
a computation would be (\ref{bigsumandintegral2}).

One ingredient in an analysis of the variation under linear
s would be the behaviour of the
variation of the laplacian acting on the Green's function
$G(y,z)$ when the arguments $y$ and $z$ are close.
Let us call the finite piece of this expression as
$z$ tends to $y$ the point-splitting regularization of
$(\delta\Delta/\Delta)$.
For surfaces
with disc topology it
is proportional to the Schwarzian
derivative of the mapping from the surface to the unit
disc\cite{Itzyksonprive}.
Let us state in passing that
the expression for the change of
the functional determinant under linear distortions
involving the Schwarzian derivative is also divergent at
corners and conical singularities.
One may attempt to regularize it by
discarding small circles around
the conical singularities and keeping the finite piece.
For e.g. triangles that gives a different answer than we
have computed for conformal s
in section~\ref{s:comp},
but one that is also incorrect if compared with
the exact results\cite{Itzyksonprive,Aurellunpub}.

It seems reasonable to expect that the finite in $\epsilon$
contribution in (\ref{bigsumandintegral2}) will be the same
as that for point-splitting, if we look at points
far from the boundary
and from conical singularities.
The correct contributions from boundaries, corners and conical
singularities remain to be computed.
We hope to return to these questions in the future.

We now turn to the Polyakov action for random surfaces,
expressed as a
the double functional integral over
$x^{\mu}$, embeddings in $d$--dimensional
external space,  and
internal two--dimensional metric
$g^{ab}$:
\begin{equation}
\label{Polyakovint}
Z \sim \int D[g^{ab}] D[x^{\mu}] \,
e^{-{1\over 2}[\int \sqrt{g} g^{ab} \partial_{a} x^{\mu}
\partial_{b} x^{\mu}]}.
\end{equation}
If we perform the integration over $x^{\mu}$ first,
we could formally write
\begin{equation}
\label{PolyakovintZ}
Z\, ``\sim''\, \int D[g^{ab}]\,
e^{{D\over 2}Z_{{\cal M}(g)}'(0)},
\end{equation}
where ${\cal M}$ is the surface that corresponds to the
metric $g$.
In fact,
the functional integral over metrics in
(\ref{Polyakovint}) vastly overcounts the
number of inequivalent surfaces, so
(\ref{PolyakovintZ})
is meaningless as it stands.
In the continuum theory this problem is solved by
fixing a gauge for the reparametrization invariance,
and computing the associated
Fadeev-Popov determinants\cite{Polyakov1,Alvarez},
which turn out to be proportional to the
determinants of the laplacian.
The basic problem of the resulting theory
seems to be that if the embedding dimension is
sufficiently large, these surfaces tend to degenerate
into the phase of branched polymers\cite{KKM,DHK}.

If one considers from the beginning summation over
simplicial complexes the redundancy of the metric is
almost completely removed.
If higher genus surfaces are related by a conformal distortion,
they are generally so in just one way.
There would then be no redundancy.
For the disc and the sphere,
the representations in terms
of charges used in sections~\ref{s:discDirichlet},
\ref{s:discNeumann} and \ref{s:sphere}
are undetermined up to a group of Moebius transformations,
which depends on six real parameters on the sphere,
and on three real parameters on the disc.
This invariance group is much smaller compared to
the intergral over metrics.

Because the elimination of the reparametrization invariance
happens in a different way than in the continuum theory,
and because the expressions of the functional determinants
in simplicial complexes are not trivial extrapolations of
the results from smooth surfaces,
it is interesting to see if one can say something
about $-{D\over 2}Z_{\cal M}'(0)$ as a possible
action for random simplicial complexes. It is clear
that the ensemble of simplicial complexes
with a fixed number of corners
and conical singularies, some of which may of course
have zero exterior
angle, can be considered as a gas of interacting charges.
We are to sum over the positions and the strength of the
charges. In the end we should then go to a canonical ensemble
and sum over all possible number of charges.

According to (\ref{zprimexpansion})
the self-energies asymptotically damp large opening angles
but {\it enhance} small opening angles
as
\begin{equation}
-{D\over 2}Z_{\alpha}'(0)\sim -0.095496 {D\over{\alpha}}.
\end{equation}
If we look at how the signs in front of the logarithms
go in
(\ref{integratedZSphere}) or
(\ref{integratedZ}), we see that like charges attract and
unlike repel. In addition, positive charges attract
strongly, and negative charges weakly.
Taken together, the self-energy and the interaction energy
enhance singularities with small opening angles, and
such singularities attract each other, the stronger
the smaller the opening angles are.
For positive $D$
this alone would certainly favour the formation of many
spikes in the surface, e.g. something qualitatively like
a branched polymer.

There are however three more effects. First, for the topology
of a sphere or a disc, there are the gauge fixing conditions.
These should take away six real degrees of freedom of motion of the
charges on the Riemann sphere, and three real degrees of freedom
on the disc. We expect that they can therefore be written as ghost
charges, three on the Riemann sphere, and for the disc one in the
interior and one on the boundary. In all they compensate for the
fact that the gauge group of Moebius transformations
is not compact, but they should not matter
much for the behaviour when two charges come close.
A second effect is that we must choose an integration measure over
the positions and the strengths of the charges.
In the continuum theory the usual choice is\cite{Polyakov1,Alvarez}
\begin{equation}
\label{Polyakovmeasure}
|| \delta\sigma || =
\int_{\hat {\cal M}} \sqrt{\hat g} e^{2\sigma} (\delta\sigma)^2 d^2z.
\end{equation}
If we translate
(\ref{Polyakovmeasure}) to e.g. Schwarz-Christoffel transformations
of the plane to a simplicial complex with the toplogy of a sphere
({\it see} (\ref{scSphere}))
we could consider it to give one more piece of
an effective action,
depending on the
exterior angles and positions of branchpoints.

Finally we should consider fixed area of the surface
or fixed length of the boundary.
One reason for doing so is that the action in
(\ref{Polyakovint}) could in general contain also
terms $\int_{\cal M} \sqrt{g}\, d^2z$ and
$\int_{\partial{\cal M}} ds$,
that give respectively the area and the length of the boundary.
Alternatively we could say that
the zeta function regularization of the functional determinants
should be related to a discretization that introduces a cut-off
scale. The intrinsic length scale
of the surface in then no more arbitrary, and could be argued
to be determined by the length scale of the surface embedded
in external space \cite{Alvarez}.
The term from the normalized area
$Z_{\cal M}(0)\log{1\over{A({\cal M})}}$ will in any case
then effectively be
one more interaction between the conical singularities.

We can argue qualitatively that this effective interaction
favors smooth surfaces.
The functional determinants in
(\ref{integratedZSphere}) or
(\ref{integratedZ}) are unchanged if the branchpoints are
moved about with Moebius transforms, since the changes in the
interaction energies and the effective interaction from the
normalized area cancel.
If one has two singularities
much closer to each other than to the others,
one may bring them still closer by a Moebius transform.
If the two that are close are positive and the others are
negative, the dominant change of the interaction energy should then
come from the two that are pushed close.
But there is no change in the total energy.
It therefore seems likely that the effective interaction
of the normalized area evens out with the attractive interaction
of like positive charges.

The normalized areas also counteracts the self-energies.
Assume some local parameter space ($|u|\le r$) for a conical
singularity,
with some small radius $r$.
This local parameter space would be the a small circular disc
around a branchpoint
for surfaces
with disc toplogy.
Conical singularities with different opening angles will be mapped from $u$
by $z\sim {1\over{\alpha}}u^{\alpha}$, and then have areas
$A(\alpha,r)\sim  r^{2\alpha}/{\alpha}$.
This would give
\begin{eqnarray}
\label{areadamping}
-Z_{\alpha}(0)\log A(\alpha,r) \sim & -{1\over{12\alpha}}\log{1\over{\alpha}}
-{1\over{12}}\log r^2
\qquad &(\alpha\, \hbox{small}) \\
-Z_{\alpha}(0)\log A(\alpha,r) \sim &
-{1\over{12}}\alpha\log\alpha
+ {1\over{12}}\alpha^2\log r^2
\qquad &( \alpha\, \hbox{large})
\end{eqnarray}
Hence here both large and small angle
corners are damped, and small angles stronger damped
than the enhancement from the self-energy.
The argument from (\ref{areadamping}) is not conclusive
of course. It assumes that one may take out one part
of the normalized surface, while the effective interaction
is proportional to the logarithm of the area of the
surface as a whole.
One could imagine a surface
with a lot of both positive and negative charges such that
the total area is around unity. It would then not
be damped by the simple effect considered in (\ref{areadamping}).

Needless to say much work remains to be done to see
whether $-{D\over 2}Z_{\cal M}'(0)$ is
an interesting action for random simplicial complexes.
Perhaps the question would
have to be answered by a numerical investigation.

\section{Acknowledgements}
E.A. thanks Ivan Kostov for valuable remarks.
This work was supported by the
Swedish Natural Science Research Council under contracts
S--FO~1778--302 and F-FU 8230-306.

\appendix
\section{Summary of properties of $Z_{\alpha}'(0)$}
\label{aexpan}
In this appendix we summarize formulae from I
concerning $Z_{\alpha}'(0)$, the corner
self-energy contribution
to the functional determinant.
The formula written in this appendix are for
a corner on the boundary with Dirichlet boundary conditions
and opening angle $\pi\alpha$.
For Neumann boundary conditions we get instead
$Z_{\alpha}'(0)+{1\over 2}\log\alpha$.
For a conical singularity in the interior with
opening angle $2\pi\alpha$ we get instead
$2Z_{\alpha}'(0)+{1\over 2}\log\alpha$.
$Z_{\alpha}'(0)$ can be written as an integral in
different ways, of which one is
\begin{eqnarray}
\label{cornerprime16}
Z'_{\alpha}(0) = {1\over{12}}({1\over{\alpha}} -\alpha)(\gamma -\log 2)
-{1\over{12}}({1\over{\alpha}} + 3 +\alpha)\log\alpha +\tilde J(\alpha),
\end{eqnarray}
\begin{eqnarray}
\label{cornerprime17}
\tilde J(\alpha) =
\int_{0}^{\infty}dy
{1\over{e^{y}-1}}
[({1\over {2y}})(\coth({{y}\over{2\alpha}})
- \alpha\coth({{y}\over{2}})) -
{1\over{12}}({1\over{\alpha}}-\alpha)].
\end{eqnarray}
For small angles the following asymptotic expansion holds:
\begin{eqnarray}
\label{zprimexpansion}
Z_{\alpha}'(0) &\sim_{\alpha\to 0}&
{1\over{\alpha}}({1\over{12}}(1-\log 2)-\zeta'(-1))
+ \alpha ({{\gamma+\log 2}\over{12}}-{1\over 4}\log 2\pi-\zeta'(-1))
\nonumber \\
&\, & +\sum_{n=3}^{\infty} {{\zeta(n)B_{n+1}}\over{n(n+1)}}\alpha^n.
\end{eqnarray}
The leading behaviour in (\ref{zprimexpansion}) is
$(0.190992\ldots)/\alpha$.
The asymptotic expansion for large angles reads:
\begin{eqnarray}
Z_{\alpha}'(0) &\sim_{\alpha\to \infty}&
-{1\over{12}}({1\over{\alpha}} +3+\alpha)\log\alpha +
\alpha ({1\over{12}}(1-\log 2)-{1\over 4}\log 2\pi-2\zeta'(-1))
\nonumber \\
&\, & + {1\over{\alpha}} ({{\gamma-\log 2}\over{12}})
+\sum_{n=3}^{\infty} {{\zeta(n)B_{n+1}}\over{n(n+1)}}\alpha^{-n},
\end{eqnarray}
for which the leading behaviour is $-{1\over {12}}\alpha\log\alpha$.
We may also expand around a flat surface i.e. $\alpha$ close to one:
\begin{eqnarray}
\label{zprimexpb}
Z_{1+\epsilon}'(0) &=&
({1\over 6}\log 2 - {5\over{24}} -
{1\over 4}\log(2\pi) - \zeta'(-1))\epsilon \nonumber \\
&\quad & + ({14\over 72} +{{\gamma-\log 2}\over{12}})\epsilon^2
+ (-{29\over 144} - {{\gamma-\log 2}\over{12}})\epsilon^3
+ {\cal O}(\epsilon^4).
\end{eqnarray}
For rational $\alpha$ we evaluated in I in closed form
$Z_{\alpha}'(0)$ to be
\begin{eqnarray}
\label{zprimepqcomplete}
Z'_{p/q}(0)={{q-p}\over{4q}}\log(2\pi)+{{p^2-q^2}\over{12pq}}\log(2)-
{1\over q}\left(p-{1\over p}\right)\zeta'(-1)-\nonumber\\
{1\over{12pq}}\log(q)+
\left({1\over4}+S(q,p)\right)\log({q\over p})+ \label{Zpq}\\
\sum_{r=1}^{p-1}\left({1\over2}-{r\over p}\right)
\log \Gamma({R(rq,p)\over p})+\sum_{s=1}^{q-1}
\left({1\over2}-{s\over q}\right)\log \Gamma({R(sp,q)\over q}),
\nonumber
\end{eqnarray}
where $R(p,q)$ and $S(p,q)$ are defined by:
\begin{eqnarray}
\label{RSdef}
R(q,p)&\equiv& q-p\left[{q\over p}\right], \\
S(q,p)&\equiv&
{1\over p}\sum_{r=0}^{p-1}r\left({R(rq,p)\over p}-{1\over2}\right).
\end{eqnarray}
For the special case $p=1$, (\ref{zprimepqcomplete})
reduces to:
\begin{eqnarray}
Z'_{1/n}(0)={1\over 4}\left(1-{1\over n}\right)\log(\pi)-
\left({n\over{12}}-{1\over 4}+{1\over{6n}}\right)\log(2)+ \nonumber\\
\left({1\over 4}-{1\over{12n}}\right)\log(n)+
\sum_{\nu=1}^{n-1}\left({1\over 2}-{\nu\over n}\right)
\log\left(\Gamma({\nu\over n})\right).\label{Z1n}
\end{eqnarray}
In the test with integrable domains we need
$Z'_{1/n}(0)$ for $n=2,3,4,6$ which are therefore reproduced
in the table below.
\bigskip

\begin{tabular}{|l|l|}
\hline
$n$ & $Z'_{1/n}(0)$ \\ \hline
2 & $\log \pi^{1\over 8} 2^{5\over 24}$	\\ \hline
3 & $\log {{\pi^{1\over 6} 2^{-{1\over 18}} 3^{4\over 9} \Gamma^{1\over
6}({1\over 3})}\over{\Gamma^{1\over 6}({2\over 3})}}$\\ \hline
4 & $\log {{\pi^{3\over 16} 2^{1\over 3} \Gamma^{1\over 4}({1\over
4})}\over{\Gamma^{1\over 4}({3\over 4})}}$	\\ \hline
6 & $\log {{\pi^{5\over 24}
2^{5\over 72} 3^{17\over 72}
\Gamma^{5\over 6}({1\over 3})}\over{\Gamma^{5\over 6}({2\over 3})}}$	\\	\hline
\end{tabular}

\section{Exactly solvable spectra}
\label{asolvable}
In this appendix we list the orbifolds and planar polygons that
can be obtained by quoting a
torus with a symmetry. The zeta functions
are computed from the spectrum with methods from number theory.
See I for a computation of cases (\ref{equilateralzetaDirichlet}),
(\ref{halfsquarezetaDirichlet}) or
(\ref{halfequilateralzetaDirichlet}) below, or \cite{Itz} for a recent
review.
\\\\
{\it Rhomboidal lattice:}
\\\\
A rhomboidal lattice
$\Lambda^{\tau}$ is spanned by two basis vectors
$\vec{\hbox{\bf e}}_1$
and $\vec{\hbox{\bf e}}_2$. We may without restriction take
$\vec{\hbox{\bf e}}_1$ to be the unit vector. The two-dimensional
plane quoted by the lattice is then identified as the torus with modular
parameter $\tau$ equal to $\vec{\hbox{\bf e}}_2$, read as a complex number.

The eigenfunctions of the laplacian on the torus are
\begin{equation}
\label{rhomboidaleigenfunctions}
\Psi^{\tau}_{mn} = e^{2\pi i (m\vec{\hbox{\bf f}}_1 +n
\vec{\hbox{\bf f}}_2)\cdot \vec x}
\qquad m\in Z, n\in Z
\end{equation}
where $\vec{\hbox{\bf f}}_1$ and $\vec{\hbox{\bf f}}_2$
are the two dual basis vectors.
The basis and the dual basis satisfy
$\vec{\hbox{\bf e}}_i\cdot \vec{\hbox{\bf f}}_j = \delta_{ij}$.
The eigenvalues of the laplacian and the spectral zeta function are
\begin{eqnarray}
\label{rhomboidaleigenvalues}
\lambda_{mn} &=& 4\pi^2 (m^2\, |\vec{\hbox{\bf f}}_1|^2+
2mn\,\vec{\hbox{\bf f}}_1\cdot \vec{\hbox{\bf f}}_2 +
n^2\, |\vec{\hbox{\bf f}}_2|^2) \\
\label{zetalambdaq}
Z(\Lambda^q,s) &=& \sum_{(m.n)\neq (0,0)}
(m^2\, |\vec{\hbox{\bf f}}_1|^2+
2mn\,\vec{\hbox{\bf f}}_1\cdot \vec{\hbox{\bf f}}_2 +
n^2\, |\vec{\hbox{\bf f}}_2|^2)^{-s}
\end{eqnarray}
For simplicity
we have in this appendix
chosen to form the zeta function without the overall
prefactor $4\pi^2$ in the eigenvalues.

The behaviour around the origin of (\ref{zetalambdaq}) is
\cite{ItzyksonZuber}
\begin{eqnarray}
Z(\Lambda^{\tau},0) = -1
&\quad &
Z'(\Lambda^{\tau},0) = -{1\over 4}\log\hbox{Area} -
\log \tau^{1\over 4}\eta(q)  \\
q=e^{2\pi i\tau}
&\quad &
\eta(q) = q^{1\over {24}}\prod_{m=1}^{\infty}(1-q^m)
\end{eqnarray}
where $\hbox{Area}$ is the area of the torus, and $\eta$ is
the modular form of Dedekind.

In general the only subgroup of the rhomboidal lattice is
a rotation through $\pi$. Quoting out with that symmetry leads to
rhombiodal envelope, visualized as two smaller copies of the
torus with free sides, glued together at the four sides.
The topology of the envelope is the sphere. The exterior angles
at the corners are all $\pi$.
The eigenvalues of the laplacian on the envelope are the
same as on the torus, except that the degeneracy between
states $(m,n)$ and $(-m,-n)$ have been divided out.
\vspace{0.5cm}

{\it Rhombic lattice:}
\\\\
A rhombic lattice is generated by two lattice vectors of the
same length. That means that the modular parameter is of
absolute value one.
In addition to rotation through $\pi$ there are now two
reflexion symmetries, that we call $S_1$ and
$S_2$. The reflexion symmetry
lines must necessarily be normal to each other.
When we quote with a reflexion symmetry, we must either add
(for Neumann boundary conditions) or subtract (for Dirichlet
boundary conditions) points on the reflexion line in the dual
lattice.
We therefore need two auxillary number sequences:
\begin{eqnarray}
l_1 &=&
\{|\vec{\hbox{\bf f}}_1+\vec{\hbox{\bf f}}_2|^2 m^2;\, m>0 \}\\
l_2 &=&
\{ |\vec{\hbox{\bf f}}_1-\vec{\hbox{\bf f}}_2|^2 m^2;\, m>0 \}
\end{eqnarray}
for which the spectral quantities are expressed in terms of
the Riemann zeta function:
\begin{eqnarray}
\label{zetal1}
Z(l_1,s) &=& \sum_{m>0}
|\vec{\hbox{\bf f}}_1+\vec{\hbox{\bf f}}_2|^{-2s}(m^2)^{-s} = |\vec{\hbox{\bf
f}}_1+\vec{\hbox{\bf f}}_2|^{-2s} \zeta(2s) \\
Z(l_2,s) &=& \sum_{m>0}
|\vec{\hbox{\bf f}}_1-\vec{\hbox{\bf f}}_2|^{-2s}(m^2)^{-s} = |\vec{\hbox{\bf
f}}_1-\vec{\hbox{\bf f}}_2|^{-2s} \zeta(2s)
\end{eqnarray}

The subgroups, their associated
surfaces and spectral quantities are
\begin{center}
\begin{enumerate}
\item $H=\{ 1 \}$. This is rhombic torus, spectrum $\Lambda_1^{\tau}$.

\item $H=\{ 1, R_{\pi} \}$. This is a rhombic envelope, spectrum
$\Lambda_{1\over 2}^{\tau}$.

\item $H=\{ 1, S_1 \}$ or $H=\{ 1, S_2 \}$ These are
both Moebius bands with spectra respectively
$\Lambda_{1\over 2}^{\tau} \pm l_1$
and
$\Lambda_{1\over 2}^{\tau} \pm l_2$.

\item $H=\{ 1, S_1, S_2, R_{\pi} \}$.
This is a cone with opening angle $\pi$. The base has two lines
and two corners, both with opening angle $\pi/2$.
The topology is that of a disc.
The spectra are
${1\over 2}(\Lambda_{1\over 2}^{\tau} \pm (l_1+l_2))$.
\end{enumerate}
\end{center}
\vspace{0.5cm}
{\it Rectangular lattice:}
\\\\
A rectangular lattice is generated by two lattice vectors normal
to each other. That means that the modular parameter is
purely imaginary.
The symmetries are the reflexions in the two lines
parallel to the basis vectors
and a rotation through $\pi$.
It is again convenient to introduce number sequences
$l_1$ and $l_2$, but this time defined as
\begin{eqnarray}
l_1 &=&
\{|\vec{\hbox{\bf f}}_1|^2 m^2;\, m>0 \}\\
l_2 &=&
\{ |\vec{\hbox{\bf f}}_2|^2 m^2;\, m>0 \}
\end{eqnarray}

The subgroups, their associated
surfaces and spectral quantities are
\begin{center}
\begin{enumerate}
\item $H=\{ 1 \}$. This is rectangular torus, spectrum $\Lambda^{\tau}_1$.

\item $H=\{ 1, R_{\pi} \}$. This is a rectangular envelope, spectrum
$\Lambda^{\tau}_{1\over 2}$.

\item $H=\{ 1, S_0 \}$ or $H=\{ 1, S_{{\pi}\over 2} \}$ These are
rectangular bands. Spectra are
$\Lambda^{\tau}_{1\over 2} \pm l_1$ and $\Lambda^{\tau}_{1\over 2} \pm l_2$.

\item $H= \{ 1, S_0, S_1, R_{\pi} \}$.
This is a rectangle. Sidelength is half of the initial
rectangular torus. Spectra are
${1\over 2}(\Lambda^{\tau}_{1\over 2} \pm (l_1+l_2))$.

\end{enumerate}
\end{center}
\vspace{0.5cm}
{\it Hexagonal lattice:}
\\\\
The two-dimensional hexagonal lattice is spanned by the
basis
\begin{equation}
\label{hexagonalroots}
\vec{\hbox{\bf e}}_1 = (1,0) \qquad
\vec{\hbox{\bf e}}_2 = (-{1\over 2},{{\sqrt{3}}\over 2})
\end{equation}
The quotient of the plane with this lattice is identified
with the torus with modular parameter $\tau = e^{\scriptscriptstyle {{2\pi
i}\over 3}}$.
The dual basis to (\ref{hexagonalroots}) is
\begin{equation}
\label{hexagonaldualroots}
\vec{\hbox{\bf f}}_1 = (1,{1\over{\sqrt{3}}}) \qquad
\vec{\hbox{\bf f}}_2 = (0,{2\over{\sqrt{3}}})
\end{equation}
The eigenvalues of the laplacian are
\begin{equation}
\label{hexagonaleigenvalues}
\lambda_{mn} = 4\pi^2 {4\over 3} (m^2 + mn + n^2)
\end{equation}
To simplify  we will in the following ignore the
overall prefactor $4\pi^2 {4\over 3}$ in the eigenvalues.

The Weyl group of the hexagonal lattice consists
of six rotations and
six reflexions. Call $R_{\alpha}$ the rotation through angle an
$\alpha$, which must be a multiple of $\pi/3$,
and $S_{\beta}$ the reflexion through a line inclined
an angle $\beta$ to the horizontal.
By convention we take take $\beta$'s equal to $\nu\pi/6$,
with $\nu=0,\ldots ,5$.
The symmetry lines of even $\nu$ are parallel to vectors in the hexagonal
lattice, while the lines of odd $\nu$ are parallel to vectors in the
dual lattice.
They hence transform the lattice differently.
The subgroups of the Weyl group
can contain either
only rotations,
or as many reflexions as rotations.
Since reflexions with even and odd $\nu$ act differently there
are two inequivalent subgroups containing either one or three reflexions.
In all we have ten inequivalent surfaces. Counting Neumann and Dirichlet
and boundary conditions for the surfaces with a boundary
(symmetric and anti-symmetric representations
of the reflexions in the
subgroup) we arrive at 16 different spectra.

It is convenient
to first introduce a notation for some spectral sequences:
\begin{eqnarray}
\Lambda_1^{e^{\scriptscriptstyle {{2\pi i}\over 3}}} &=& \{ m^2 +mn +n^2; m\in
Z, n\in Z; (m,n) \neq (0,0) \} \\
\Lambda_{1\over 2}^{e^{\scriptscriptstyle {{2\pi i}\over 3}}} &=&
\{\Lambda_1^{e^{\scriptscriptstyle {{2\pi i}\over 3}}}; m>0,\, \hbox{or}\, m=0,
n>0 \}\\
\Lambda_{1\over 3}^{e^{\scriptscriptstyle {{2\pi i}\over 3}}} &=&
\{\Lambda_1^{e^{\scriptscriptstyle {{2\pi i}\over 3}}}; m>0, n\le 0 \}\\
\Lambda_{1\over 6}^{e^{\scriptscriptstyle {{2\pi i}\over 3}}} &=&
\{\Lambda_1^{e^{\scriptscriptstyle {{2\pi i}\over 3}}}; m>0, n\ge 0 \}\\
l_1 &=& \{m^2; m>0 \}\\
l_{1\over{\sqrt{3}}} &=& \{3m^2; m>0 \}
\end{eqnarray}
The meaning of these sequences is that $\Lambda_k^{e^{\scriptscriptstyle {{2\pi
i}\over 3}}}$
is the lattice quoted by a subgroup of order $k$ containing
only rotations.
The spectral functions
will be expressed in terms of the Riemann zeta function and
\begin{eqnarray}
\label{zetalambda1over6}
Z(\Lambda_{1\over 6}^{e^{\scriptscriptstyle {{2\pi i}\over 3}}},s) =
\sum_{m>0,n\ge 0} (m^2+mn+n^2)^{-s}
&=& L_3(s)\zeta(s)
\end{eqnarray}
with
\begin{equation}
L_3(s)= 1-2^{-s}+4^{-s}-5^{-s}+\cdots
\end{equation}
The behaviour around the origin of the Riemann zeta function
and
(\ref{zetalambda1over6})
is
\begin{eqnarray}
\zeta(0) = -{1\over 2} &\quad& \zeta'(0) = -{1\over 2}\log 2\pi \\
Z(\Lambda_{1\over 6}^{e^{\scriptscriptstyle {{2\pi i}\over 3}}},0) = -{1\over
6}
&\quad& Z'(\Lambda_{1\over 6}^{e^{\scriptscriptstyle {{2\pi i}\over 3}}},0) =
{1\over 2}\log{{\Gamma({2\over 3})}\over {\Gamma({1\over 3})}}
- {1\over 6}\log{{2\pi}\over 3}
\end{eqnarray}

We now list
the possible inequivalent subgroups, their associated
surfaces and spectral quantities.
For surfaces with boundary we use the convention
that the result for Neumann boundary conditions is given first
(generally with a plus sign), and then the result with Dirichlet
boundary conditions (generally with a minus sign).
\begin{center}
\begin{enumerate}
\item $H=\{ 1 \}$. This is the torus itself, spectrum
$\Lambda_1^{e^{\scriptscriptstyle {{2\pi i}\over 3}}}$.
\begin{eqnarray}
Z(\Lambda_1^{e^{\scriptscriptstyle {{2\pi i}\over 3}}},0) = -1 &\quad&
Z'(\Lambda_1^{e^{\scriptscriptstyle {{2\pi i}\over 3}}},0) =
3\log{{\Gamma({2\over 3})}\over {\Gamma({1\over 3})}}
- \log{{2\pi}\over 3}
\end{eqnarray}

\item $H=\{ 1, R_{\pi} \}$. This is the tetrahedron, spectrum
$\Lambda_{1\over 2}^{e^{\scriptscriptstyle {{2\pi i}\over 3}}}$.
\begin{eqnarray}
Z(\Lambda_{1\over 2}^{e^{\scriptscriptstyle {{2\pi i}\over 3}}},0) = -{1\over
2} &\quad& Z'(\Lambda_{1\over 2}^{e^{\scriptscriptstyle {{2\pi i}\over 3}}},0)
=
{3\over 2}\log{{\Gamma({2\over 3})}\over {\Gamma({1\over 3})}}
- {1\over 2}\log{{2\pi}\over 3}
\end{eqnarray}

\item $H=\{ 1, R_{\pm {{2\pi}\over 3}} \}$. This is the equilateral triangle
envelope,
visualized as two equilateral triangels glued together at the three sides.
Spectrum is $\Lambda_{1\over 3}^{e^{\scriptscriptstyle {{2\pi i}\over 3}}}$.
\begin{eqnarray}
Z(\Lambda_{1\over 3}^{e^{\scriptscriptstyle {{2\pi i}\over 3}}},0) = -{1\over
3} &\quad& Z'(\Lambda_{1\over 3}^{e^{\scriptscriptstyle {{2\pi i}\over 3}}},0)
=
\log{{\Gamma({2\over 3})}\over {\Gamma({1\over 3})}}
- {1\over 3}\log{{2\pi}\over 3}
\end{eqnarray}

\item $H=\{ 1,R_{\pm {{\pi}\over 3}}, R_{\pm {{2\pi}\over 3}},
R_{\pi} \}$. This is bisected equilateral triangle envelope,
visualized as two triangels with angles $\pi/2$, $\pi/3$ and $\pi/6$,
glued together at the three sides.
Spectrum is $\Lambda_{1\over 6}^{e^{\scriptscriptstyle {{2\pi i}\over 3}}}$.
\begin{eqnarray}
Z(\Lambda_{1\over 6}^{e^{\scriptscriptstyle {{2\pi i}\over 3}}},0) = -{1\over
6} &\quad& Z'(\Lambda_{1\over 6}^{e^{\scriptscriptstyle {{2\pi i}\over 3}}},0)
=
{1\over 2}\log{{\Gamma({2\over 3})}\over {\Gamma({1\over 3})}}
- {1\over 6}\log{{2\pi}\over 3}
\end{eqnarray}

\item $H=\{ 1, S_{{\pi}\over 3} \}$. This is a Moebius band.
The spectra are
$\Lambda_{1\over 2}^{e^{\scriptscriptstyle {{2\pi i}\over 3}}}\pm
l_{1\over{\sqrt{3}}}$.
\begin{eqnarray}
Z(\Lambda_{1\over 2}^{e^{\scriptscriptstyle {{2\pi i}\over
3}}}\!+\!l_{1\over{\sqrt{3}}},0) = -1 &\,&
Z'(\Lambda_{1\over 2}^{e^{\scriptscriptstyle {{2\pi i}\over
3}}}\!+\!l_{1\over{\sqrt{3}}},0) =
{3\over 2}\log{{\Gamma({2\over 3})}\over {\Gamma({1\over 3})}}
+\log 3 - {3\over 2}\log 2\pi \\
Z(\Lambda_{1\over 2}^{e^{\scriptscriptstyle {{2\pi i}\over
3}}}\!-\!l_{1\over{\sqrt{3}}},0) = 0 &\,&
Z'(\Lambda_{1\over 2}^{e^{\scriptscriptstyle {{2\pi i}\over
3}}}\!-\!l_{1\over{\sqrt{3}}},0) =
{3\over 2}\log{{\Gamma({2\over 3})}\over {\Gamma({1\over 3})}}
+{1\over 2}\log 2\pi
\end{eqnarray}

\item $H=\{ 1, S_{{5\pi}\over 6} \}$. This is another Moebius band.
The spectra are $\Lambda_{1\over 2}^{e^{\scriptscriptstyle {{2\pi i}\over
3}}}\pm l_1$.
\begin{eqnarray}
Z(\Lambda_{1\over 2}^{e^{\scriptscriptstyle {{2\pi i}\over 3}}}\!+\!l_1,0) = -1
&\,&
Z'(\Lambda_{1\over 2}^{e^{\scriptscriptstyle {{2\pi i}\over 3}}}\!+\!l_1,0) =
{3\over 2}\log{{\Gamma({2\over 3})}\over {\Gamma({1\over 3})}}
+{1\over 2}\log 3 - {3\over 2}\log 2\pi \\
Z(\Lambda_{1\over 2}^{e^{\scriptscriptstyle {{2\pi i}\over 3}}}\!-\!l_1,0) = 0
&\,&
Z'(\Lambda_{1\over 2}^{e^{\scriptscriptstyle {{2\pi i}\over 3}}}\!-\!l_1,0) =
{3\over 2}\log{{\Gamma({2\over 3})}\over {\Gamma({1\over 3})}}
+{1\over 2}\log 3+{1\over 2}\log 2\pi
\end{eqnarray}

\item $H=\{ 1, S_{{\pi}\over 3}, S_{{5\pi}\over 6}, R_{\pi} \}$.
This is a cone with opening angle
$\pi$. The base has one longer and one shorter
side, and two corners with angles $\pi/2$.
The spectra are
${3\over 2}\Lambda_{1\over 6}^{e^{\scriptscriptstyle {{2\pi i}\over 3}}}\pm
(l_1+l_{1\over{\sqrt{3}}})$.
\begin{eqnarray}
Z({3\over 2}\Lambda_{1\over 6}^{e^{\scriptscriptstyle {{2\pi i}\over
3}}}\!+\!l_1\!+\!l_{1\over{\sqrt{3}}},0) = -{3\over 4} &\,&
Z'({3\over 2}\Lambda_{1\over 6}^{e^{\scriptscriptstyle {{2\pi i}\over
3}}}\!+\!l_1\!+\!l_{1\over{\sqrt{3}}},0) =
{3\over 4}\log{{\Gamma({2\over 3})}\over {\Gamma({1\over 3})}}
\nonumber \\ & & \qquad +{1\over 2}\log 3-{3\over 2}\log 2\pi \\
Z({3\over 2}\Lambda_{1\over 6}^{e^{\scriptscriptstyle {{2\pi i}\over 3}}}\!-\!
l_1\!-\!l_{1\over{\sqrt{3}}},0) = {1\over 4} &\,&
Z'({3\over 2}\Lambda_{1\over 6}^{e^{\scriptscriptstyle {{2\pi i}\over
3}}}\!-\!l_1\!-\!l_{1\over{\sqrt{3}}},0) =
{3\over 4}\log{{\Gamma({2\over 3})}\over {\Gamma({1\over 3})}}
\nonumber \\ & & \qquad + {1\over 2}\log 2\pi
\end{eqnarray}

\item $H=\{ 1, S_0, S_{{\pi}\over 3}, S_{{2\pi}\over 3},
 R_{\pm {{2\pi}\over 3}} \}$.
This is a cone with opening angle
${{2\pi}\over 3}$. The base has one side and one corner
with angle ${{\pi}\over 3}$.
The spectra are
$\Lambda_{1\over 6}^{e^{\scriptscriptstyle {{2\pi i}\over 3}}}\pm
l_{1\over{\sqrt{3}}}$.
\begin{eqnarray}
Z(\Lambda_{1\over 6}^{e^{\scriptscriptstyle {{2\pi i}\over
3}}}\!+\!l_{1\over{\sqrt{3}}}),0) = -{2\over 3} &\,&
Z'(\Lambda_{1\over 6}^{e^{\scriptscriptstyle {{2\pi i}\over
3}}}\!+\!l_{1\over{\sqrt{3}}}),0) =
{1\over 2}\log{{\Gamma({2\over 3})}\over {\Gamma({1\over 3})}}
\nonumber \\ & & \qquad +{2\over 3}\log 3-{7\over 6}\log 2\pi \\
Z(\Lambda_{1\over 6}^{e^{\scriptscriptstyle {{2\pi i}\over
3}}}\!-\!l_{1\over{\sqrt{3}}},0) = {1\over 3} &\,&
Z'(\Lambda_{1\over 6}^{e^{\scriptscriptstyle {{2\pi i}\over
3}}}\!-\!l_{1\over{\sqrt{3}}},0) =
{1\over 2}\log{{\Gamma({2\over 3})}\over {\Gamma({1\over 3})}}
\nonumber \\ & & \qquad -{1\over 3}\log 3+{5\over 6}\log 2\pi
\end{eqnarray}

\item $H=\{ 1, S_{{\pi}\over 6}, S_{{\pi}\over 2}, S_{{5\pi}\over 6},
R_{\pm {{2\pi}\over 3}} \}$.
This is the equilateral triangle.
The spectra are
$\Lambda_{1\over 6}^{e^{\scriptscriptstyle {{2\pi i}\over 3}}}\pm l_1$.

\begin{eqnarray}
Z(\Lambda_{1\over 6}^{e^{\scriptscriptstyle {{2\pi i}\over 3}}}\!+\!l_1),0) =
-{2\over 3} &\,&
Z'(\Lambda_{1\over 6}^{e^{\scriptscriptstyle {{2\pi i}\over 3}}}\!+\!l_1),0) =
{1\over 2}\log{{\Gamma({2\over 3})}\over {\Gamma({1\over 3})}}
\nonumber \\ & & \qquad +{1\over 6}\log 3 - {7\over 6}\log 2\pi \\
\label{equilateralzetaDirichlet}
Z(\Lambda_{1\over 6}^{e^{\scriptscriptstyle {{2\pi i}\over 3}}}\!-\!l_1,0) =
{1\over 3} &\,&
Z'(\Lambda_{1\over 6}^{e^{\scriptscriptstyle {{2\pi i}\over 3}}}\!-\!l_1,0) =
{1\over 2}\log{{\Gamma({2\over 3})}\over {\Gamma({1\over 3})}}
\nonumber \\ & & \qquad - {1\over 6}\log 3
+ {5\over 6}\log 2\pi
\end{eqnarray}

\item $H=\hbox{Entire Weyl group}$.
This is the bisected equilateral triangle
with angles
$\pi/2$, $\pi/3$ and $\pi/6$.
The spectra are
$\frac{1}{2}(\Lambda_{1\over 6}^{e^{\scriptscriptstyle {{2\pi i}\over 3}}}\pm
(l_1+l_{1\over{\sqrt{3}}}))$.

\begin{eqnarray}
Z(\frac{1}{2}(\Lambda_{1\over 6}^{e^{\scriptscriptstyle {{2\pi i}\over
3}}}\!+\!l_1\!+\!l_{1\over{\sqrt{3}}}),0) =
-\frac{7}{12} &\,&
Z'(\frac{1}{2}(\Lambda_{1\over 6}^{e^{\scriptscriptstyle {{2\pi i}\over
3}}}\!+\!l_1\!+\!l_{1\over{\sqrt{3}}}),0) =
\frac{1}{4}\log{{\Gamma({2\over 3})}\over {\Gamma({1\over 3})}}
\nonumber \\ & & \qquad+\frac{7}{3}\log 3-\frac{13}{12}\log 2\pi \\
\label{halfequilateralzetaDirichlet}
Z(\frac{1}{2}(\Lambda_{1\over 6}^{e^{\scriptscriptstyle {{2\pi i}\over
3}}}\!-\!l_1\!-\!l_{1\over{\sqrt{3}}}),0) =
\frac{5}{12} &\,&
Z'(\frac{1}{2}(\Lambda_{1\over 6}^{e^{\scriptscriptstyle {{2\pi i}\over
3}}}\!-\!l_1\!-\!l_{1\over{\sqrt{3}}}),0) =
\frac{1}{4}\log{{\Gamma({2\over 3})}\over {\Gamma({1\over 3})}}
\nonumber \\ & & \qquad-\frac{1}{6}\log 3+\frac{11}{12}\log 2\pi
\end{eqnarray}

\end{enumerate}
\end{center}
\vspace{0.5cm}
{\it Square lattice:}
\\\\
The eigenfunctions of the laplacian on the square
torus are
\begin{equation}
\label{squareeigenfunctions}
\Psi_{mn} = e^{2\pi i (m x_1+n x_2)}
\qquad m\in Z, n\in Z
\end{equation}
with eigenvalues
\begin{equation}
\label{squareeigenvalues}
\lambda_{mn} = 4\pi^2 (m^2 + n^2)
\end{equation}
The modular parameter is $i$.

The Weyl group of the square lattice consists
of four rotations and
four reflexions.
The reflexion
symmetry lines parallel to the diagonals
($S_{{\pi}\over 4}$ and $S_{{3\pi}\over 4}$)
act differently than reflexions in the lines
parallel to the lattice vectors
($S_0$ and $S_{{\pi}\over 2}$).
We have thus 8 inequivalent subgroups, and counting
Neumann
and Dirichlet boundary conditions for the surfaces with a boundary,
13 different spectra.

It is again convenient
to introduce a notation for some spectral sequences:
\begin{eqnarray}
\Lambda_1^{i} &=& \{ m^2 +n^2; m\in Z, n\in Z; (m,n) \neq (0,0) \} \\
\Lambda_{1\over 2}^{i} &=& \{\Lambda_1; m>0,\, \hbox{or}\, m=0, n>0 \}\\
\Lambda_{1\over 4}^{i} &=& \{\Lambda_1; m>0, n \ge 0 \}\\
l_1 &=& \{m^2; m>0 \}\\
l_{1\over{\sqrt{2}}} &=& \{2m^2; m>0 \}
\end{eqnarray}
We will need
\begin{eqnarray}
\label{zetalambda1over4}
Z(\Lambda_{1\over 4},s) = \sum_{m>0,n\ge 0} (m^2+n^2)^{-s}
&=& L_4(s)\zeta(s)
\end{eqnarray}
with
\begin{equation}
L_4(s)= 1-3^{-s}+5^{-s}-7^{-s}+\cdots
\end{equation}
The behaviour around the origin of
(\ref{zetalambda1over4})
is
\begin{eqnarray}
Z(\Lambda_{1\over 4}^{i},0) = -{1\over 4}
&\quad& Z'(\Lambda_{1\over 4}^{i},0) =
{1\over 2}\log{{\Gamma({3\over 4})}\over {\Gamma({1\over 4})}}
+{1\over 4}\log{1\over{2\pi}}  +{1\over 2}\log 2
\end{eqnarray}

We now list
the possible inequivalent subgroups, their associated
surfaces and spectral quantities.
\begin{center}
\begin{enumerate}
\item $H=\{ 1 \}$. This is the square torus, spectrum $\Lambda^{i}_1$.
\begin{eqnarray}
Z(\Lambda_1^{i},0) = -1 &\quad& Z'(\Lambda_1^{i},0) =
2\log{{\Gamma({3\over 4})}\over {\Gamma({1\over 4})}}
+\log{1\over{2\pi}}  +2\log 2
\end{eqnarray}

\item $H=\{ 1, R_{\pi} \}$. This is a square envelope, spectrum
$\Lambda_{1\over 2}^{i}$.
\begin{eqnarray}
Z(\Lambda_{1\over 2}^{i},0) = -{1\over 2} &\quad& Z'(\Lambda_{1\over 2}^{i},0)
=
\log{{\Gamma({3\over 4})}\over {\Gamma({1\over 4})}}
+{1\over 2}\log{1\over{2\pi}}  +\log 2
\end{eqnarray}

\item $H=\{ 1, R_{\pm{{\pi}\over 2}},R_{\pi} \}$. This is
a right angle isosceles triangle envelope,
visualized as two triangels with angles
${{\pi}\over 2}$, ${{\pi}\over 4}$ and ${{\pi}\over 4}$
glued together at the sides.
Spectrum is $\Lambda_{1\over 4}^{i}$.
\begin{eqnarray}
Z(\Lambda_{1\over 4}^{i},0) = -{1\over 4} &\quad& Z'(\Lambda_{1\over 4}^{i},0)
=
{1\over 2} \log{{\Gamma({3\over 4})}\over {\Gamma({1\over 4})}}
+{1\over 4}\log{1\over{2\pi}}  +{1\over 2}\log 2
\end{eqnarray}

\item $H=\{ 1, S_0 \}$. This is a rectangular band.
The spectra are
$\Lambda_{1\over 2}^{i}\pm l_1$.
\begin{eqnarray}
Z(\Lambda_{1\over 2}^{i}\!+\!l_1,0) = -1 &\,&
Z'(\Lambda_{1\over 2}^{i}\!+\!l_1,0) =
\log{{\Gamma({3\over 4})}\over {\Gamma({1\over 4})}}
\nonumber \\ & & \qquad +{3\over 2}\log{1\over{2\pi}}  +\log 2
\\
Z(\Lambda_{1\over 2}^{i}\!-\!l_1,0) = 0 &\,&
Z'(\Lambda_{1\over 2}^{i}\!-\!l_1,0) =
\log{{\Gamma({3\over 4})}\over {\Gamma({1\over 4})}}
\nonumber \\ & & \qquad
-{1\over 2}\log{1\over{2\pi}}  +\log 2
\end{eqnarray}

\item $H=\{ 1, S_{{\pi}\over 4} \}$. This is Moebius band.
The spectra are $\Lambda_{1\over 2}^{i}\pm l_{1\over{\sqrt{2}}}$.
\begin{eqnarray}
Z(\Lambda_{1\over 2}^{i}\!+\!l_{1\over{\sqrt{2}}},0) = -1 &\,&
Z'(\Lambda_{1\over 2}^{i}\!+\!l_{1\over{\sqrt{2}}},0) =
\log{{\Gamma({3\over 4})}\over {\Gamma({1\over 4})}}
\nonumber \\ & & \qquad
+{3\over 2}\log{1\over{2\pi}}  +{3\over 2}\log 2
\\
Z(\Lambda_{1\over 2}^{i}\!-\!l_{1\over{\sqrt{2}}},0) = 0 &\,&
Z'(\Lambda_{1\over 2}^{i}\!-\!l_{1\over{\sqrt{2}}},0) =
\log{{\Gamma({3\over 4})}\over {\Gamma({1\over 4})}}
\nonumber \\ & & \qquad
-{1\over 2}\log{1\over{2\pi}}  +{1\over 2}\log 2
\end{eqnarray}

\item $H=\{ 1, S_0,  S_{{\pi}\over 2}, R_{\pi} \}$.
This is the square.
The spectra, which could of course equally well have been
written down directly, are
$\Lambda_{1\over 4}^{i}\pm l_1$.
\begin{eqnarray}
Z(\Lambda_{1\over 4}^{i}\!+\!l_1\!,0) = -{3\over 4} &\,&
Z'(\Lambda_{1\over 4}^{i}\!+\!l_1,0) =
{1\over 2}\log{{\Gamma({3\over 4})}\over {\Gamma({1\over 4})}}
\nonumber \\ & & \qquad +
{5\over 4}\log{1\over{2\pi}}  +{1\over 2}\log 2
\\
Z(\Lambda_{1\over 4}^{i}\!-\!l_1,0) = {1\over 4} &\,&
Z'(\Lambda_{1\over 4}^{i}\!-\!l_1,0) =
{1\over 2}\log{{\Gamma({3\over 4})}\over {\Gamma({1\over 4})}}
\nonumber \\ & & \qquad
-{3\over 4}\log{1\over{2\pi}}  +{1\over 2}\log 2
\end{eqnarray}

\item $H=\{ 1, S_{{\pi}\over 4}, S_{{3\pi}\over 4}, R_{\pi} \}$.
This is a cone with opening angle
$\pi$. The base has two sides of equal length, and
two corners
with angles $\pi/2$.
The spectra are
$\Lambda_{1\over 4}^{i}\pm l_{1\over{\sqrt{2}}}$.
\begin{eqnarray}
Z(\Lambda_{1\over 4}^{i}\!+\!l_{1\over{\sqrt{2}}}),0) = -{3\over 4}&\,&
Z'(\Lambda_{1\over 4}^{i}\!+\!l_{1\over{\sqrt{2}}}),0) =
{1\over 2}\log{{\Gamma({3\over 4})}\over {\Gamma({1\over 4})}}
\nonumber \\ & & \qquad
{5\over 4}\log{1\over{2\pi}}  +\log 2
\\
Z(\Lambda_{1\over 4}^{i}\!-\!l_{1\over{\sqrt{2}}},0) = {1\over 4} &\,&
Z'(\Lambda_{1\over 4}^{i}\!-\!l_{1\over{\sqrt{2}}},0) =
{1\over 2}\log{{\Gamma({3\over 4})}\over {\Gamma({1\over 4})}}
\nonumber \\ & & \qquad
-{3\over 4}\log{1\over{2\pi}}
\end{eqnarray}

\item $H=\hbox{Entire Weyl group}$.
This is the right angles isosceles triangle, with angles
$\pi/2$, $\pi/4$ and $\pi/4$.
The spectra are
$\frac{1}{2}(\Lambda_{1\over 4}^{i}\pm (l_1+l_{1\over{\sqrt{2}}}))$.

\begin{eqnarray}
Z(\frac{1}{2}(\Lambda_{1\over 4}^{i}\!+\!l_1\!+\!l_{1\over{\sqrt{2}}}),0) =
-{5\over 8} &\,&
Z'(\frac{1}{2}(\Lambda_{1\over 4}^{i}\!+\!l_1\!+\!l_{1\over{\sqrt{2}}}),0) =
\frac{1}{4}\log{{\Gamma({3\over 4})}\over {\Gamma({1\over 4})}}
\nonumber \\ & & \qquad+
{9\over 8}\log{1\over{2\pi}}  +{1\over 2}\log 2
\\
\label{halfsquarezetaDirichlet}
Z(\frac{1}{2}(\Lambda_{1\over 4}^{i}\!-\!l_1\!-\!l_{1\over{\sqrt{2}}}),0) =
{3\over 8}  &\,&
Z'(\frac{1}{2}(\Lambda_{1\over 4}^{i}
\!-\!l_1\!-\!l_{1\over{\sqrt{2}}}),0) =
\frac{1}{4}\log{{\Gamma({3\over 4})}\over {\Gamma({1\over 4})}}
\nonumber \\ & & \qquad
-{7\over 8}\log{1\over{2\pi}}
\end{eqnarray}
\end{enumerate}
\end{center}

\end{document}